\documentclass[journal]{IEEEtran}
\usepackage{tikz}
\usetikzlibrary{arrows,positioning,fit,shapes,calc,decorations.pathreplacing,matrix,patterns}

\usepackage{amsmath}
\usepackage{subfigure}
\usepackage{lineno,hyperref}
\usepackage{xcolor}
\usepackage{extarrows}
\usepackage{algorithm}
\usepackage{algorithmic}
\usepackage{amssymb}
\usepackage{graphicx}
\usepackage[justification=centering]{caption}    
\usepackage{cite}

\usepackage{stfloats}
\usepackage{amsthm}
\usepackage{geometry}
\usepackage{booktabs}
\newcommand{\rew}{\textcolor{black}}

\newcommand{\ba}{\boldsymbol{a}}

\newcommand{\bc}{\boldsymbol{c}}

\newcommand{\bx}{\boldsymbol{x}}
\newcommand{\by}{\boldsymbol{y}}

\newcommand{\bw}{\boldsymbol{w}}

\newcommand{\bp}{\boldsymbol{p}}
\newcommand{\bs}{\boldsymbol{s}}
\newcommand{\bq}{\boldsymbol{q}}
\newcommand{\br}{\boldsymbol{r}}

\newcommand{\be}{\boldsymbol{e}}

\newcommand{\bX}{\boldsymbol{X}}
\newcommand{\bE}{\boldsymbol{E}}
\newcommand{\bP}{\boldsymbol{P}}
\newcommand{\bY}{\boldsymbol{Y}}
\newcommand{\bA}{\boldsymbol{A}}
\newcommand{\bB}{\boldsymbol{B}}
\newcommand{\bR}{\boldsymbol{R}}

\newcommand{\bW}{\boldsymbol{W}}
\newcommand{\bI}{\boldsymbol{I}}

\newcommand{\bC}{\boldsymbol{C}}
\newcommand{\bD}{\boldsymbol{D}}
\newcommand{\bU}{\boldsymbol{U}}
\newcommand{\bV}{\boldsymbol{V}}
\newcommand{\bQ}{\boldsymbol{Q}}

\newcommand{\bS}{\boldsymbol{S}}
\newcommand{\bG}{\boldsymbol{G}}

\newcommand{\bXi}{\boldsymbol{\Xi}}
\newcommand{\bGamma}{\boldsymbol{\Gamma}}
\newcommand{\bLambda}{\boldsymbol{\Lambda}}

\newcommand{\N}{\mathcal{N}}

\newcommand{\mR}{\mathcal{R}}

\newcommand{\Tr}{\text{Tr}}

\newcommand{\tVec}{\textrm{Vec}}
\newcommand{\herm}{^{\textrm{H}}}
\newcommand{\tra}{^{\textrm{T}}}

\newcommand{\CN}{\mathcal{CN}}
\newcommand{\MN}{\mathcal{MN}}
\newcommand{\Ga}{\mathcal{G}\mathfrak{a}}

\newcommand{\bPhi}{\boldsymbol{\Phi}}

\newcommand{\balpha}{\boldsymbol{\alpha}}
\newcommand{\bxi}{\boldsymbol{\xi}}

\newcommand{\bgamma}{\boldsymbol{\gamma}}

\usetikzlibrary{arrows,positioning,fit,shapes,calc,decorations.pathreplacing,matrix,patterns}

\newgeometry{left=1.5cm, right=1.5cm, top=1.5cm, bottom=1cm}
\tikzstyle{factornode} = [draw, fill=white, circle, inner sep=1pt, minimum size=1.4cm]
\tikzstyle{funnode} = [draw, rectangle,fill=black!100, minimum size = 8mm]

\begin{document}
\title{Integrated Near Field Sensing and Communications Using Unitary Approximate Message Passing Based Matrix Factorization}
\author{Zhengdao~Yuan, Qinghua Guo, \IEEEmembership{Senior Member, IEEE}, Yonina C. Eldar, \IEEEmembership{Fellow, IEEE}, and Yonghui Li,   \IEEEmembership{Fellow, IEEE}
	%\thanks{Z. Yuan?s work was supported by the National Natural Science Foundation of China (NSFC61571402), Postdoctoral science foundation of China (2019M652576) and Henan research project of high education (20B510005).}
	%\thanks{Corresponding author: Qinghua Guo.}
	\thanks{Part of this work will be presented in ICASSP 2024.}
	\thanks{Z. Yuan is with the Artificial Intelligence Technology Engineering Research Center, Open University of Henan,
	%and School of Information Engineering, Zhengzhou University,
	Zhengzhou 450002, China. He was with the School of Electrical, Computer and Telecommunications Engineering, University of Wollongong, Wollongong, NSW 2522, Australia (e-mail: yuan\_zhengdao@foxmail.com).}
	\thanks{Q. Guo is with the School of Electrical, Computer and Telecommunications Engineering, University of Wollongong, Wollongong, NSW 2522, Australia  (e-mail: qguo@uow.edu.au).}
	\thanks{Y. C. Eldar is with the Faculty of Mathematics and Computer Science, Weizmann Institute of Science, Rehovot 7610001, Israel (e-mail: yonina.eldar@weizmann.ac.il).}
	\thanks{Y. Li is with the School of Electrical and Information Engineering, University of Sydney, Sydney, NSW 2006, Australia (e-mail: yonghui.li@sydney.edu.au).}
}
\markboth{UAMP-MF-ISAC}
{Shell \MakeLowercase{\textit{et al.}}: Bare Demo of IEEEtran.cls for IEEE Journals}

\maketitle

\begin{abstract}
Due to the utilization of large antenna arrays at base stations (BSs) and the operations of wireless communications in high frequency bands, mobile terminals often find themselves in the near-field of the array aperture. In this work, we address the signal processing challenges of integrated near-field localization and communication  in uplink transmission of an integrated sensing and communication (ISAC) system, where the BS performs joint near-field localization %(through estimating the arrival angles and distances of mobile terminals) 
and signal detection (JNFLSD). We show that JNFLSD can be formulated as a matrix factorization (MF) problem with proper structures imposed on the factor matrices. Then, leveraging the variational inference (VI) and unitary approximate message passing (UAMP), we develop a low complexity Bayesian approach to MF, called UAMP-MF, to handle a generic MF problem. We then apply the UAMP-MF algorithm to solve the JNFLSD problem, where the factor matrix structures are fully exploited. Extensive simulation results are provided to demonstrate the superior performance of the proposed method.    
\end{abstract}


\section{Introduction}

Equipped with a large number of antennas at base stations (BSs), massive multiple-input-multiple-output (MIMO) technology has the potential to significantly enhance spectral efficiency by orders of magnitude, recognized as one of the key technologies for future generation of wireless communications \cite{Cui2023}, \cite{Lu2023}. The availability of abundant spectrum resources in the millimeterwave (mmWave) and Terahertz bands presents an attractive proposition for high-frequency communications \cite{Elayan2020}. 
%makes high-frequency communications a compelling option.
Moreover, the compact dimensions of high-frequency antennas renders them highly suitable for deploying massive MIMO systems with a large number of antennas. Consequently, high-frequency massive MIMO is widely acclaimed as a key enabler for the future of wireless communications \cite{Rappaport2019}.

Owing to the deployment of large antenna apertures at base stations (BSs) and the utilization of high-frequency bands, mobile terminals frequently find themselves within the near-field of the array aperture. This proximity to the array is due to the fact that the Rayleigh distance, which serves as the boundary between the near-field and far-field regions, is typically large \cite{Grosicki2005}, \cite{Selvan2017}. For instance, consider a uniform linear antenna array comprising 128 elements with half-wavelength spacing, operating at a carrier frequency of 30GHz. In this context, the Rayleigh distance is approximately 82 meters, rendering the near-field region a significant consideration in MIMO systems. When the receiver is positioned within the near-field region, it becomes imperative to accurately model the wavefront under the spherical wavefront assumption, in contrast to the planar wavefront in the far-field region \cite{Zhou2015,Cui2021, Wei2022}.

Recently, there has been a significant upsurge of interest in integrated sensing and communication (ISAC) in wireless networks \cite{Liu2022,NOMA_ISAC2023,Wang2023,Cui2021,Boqun2023}. Sensing has now emerged as an integral element of wireless networks, e.g., localization will be a service in wireless networks for many applications such as autonomous driving. 
In the near-field scenario, the conventional localization techniques developed based on the far-field assumption e.g., the on-grid the angle of arrival estimation method simultaneous weighted-orthogonal matching pursuit (SW-OMP) \cite{Rod2018}, and the simultaneous iterative gridless weighted orthogonal least square (SIGW-OLS) angle estimation method  \cite{Gon2021} suffer from performance loss due to model mismatch.   
%in the literature, but there is a serious spectrum leakage problem, which will lead to estimation errors when the user is not accurately located on the grid \cite{Gao2016}. To solve the leakage problem, an estimation method based on simultaneous iterative gridless weighted (SIGW-OLS) is also proposed in the literature \cite{Gon2021}. However, the direct application of the far-field model in the near-field environment leads to serious estimation errors due to the model mismatch. 
Recently, the problem of target localization and channel estimation in near-field environment has received tremendous attention, e.g., the works in \cite{Han2020,Wang2023,Grosicki2005,Zuo2018,Cui2022}. The Cramer-Rao bound of target localization in near-field was studied \cite{Grosicki2005}, and the bound subject to minimum communication rate requirement was investigated in \cite{Wang2023}. % and a suitable ISAC waveform is designed \cite{Wang2023}.
In large-scale massive MIMO systems, subarray-wise and scatterer-wise
channel estimation methods to estimate the near-field nonstationary
channel were studied in \cite{Han2020}.
By using high-order statistics and leveraging the structure of the signal covariance matrix, subspace-based algorithms \cite{Zuo2018} and high-order MUSIC algorithms \cite{Gao2016} were proposed to estimate angles of departure and distances between sources and receivers in the near-field.  
%By exploiting the structure of signal covariance matrix, high-order statistic based methods, e.g., the subspace-based algorithms \cite{Zuo2018} and high-order MUSIC algorithms \cite{Friedlander2019} have been proposed to estimate the angles of departure and distance between sources and the receiver. 
Near-field channel estimation was studied in \cite{Cui2022}, and further in \cite{Wei2022}, a near-field channel estimation algorithm (NF-SOMP) was developed to handle the case that far-field and near-field paths coexist. 
%However, an algorithm that can implement blind joint near-field localization and signal detection has not yet appeared. 
In this work, we focus on the issue of integrated near-field sensing and communications in the case of uplink transmission without pilot signals, where the BS assumes the dual role of user localization and signal detection, by performing blind joint near-field localization and signal detection (JNFLSD).

In this paper, we show that blind JNFLSD can be formulated as a matrix factorization (MF) problem with imposed structures on two factor matrices, i.e., factorizing a received signal matrix $\bY$ to the product of two factor matrices $\bA$ and $\bX$ with the consideration of noise perturbation, where the factor matrix $\bX$ is sparse and the columns of  matrix $\bA$ are subject to the structure specified by the distance-angle dependent steering vectors.  
%model $\bY= \bA \bX + \bW$, where $\bY$ is the received signal matrix, $\bW$ represents the noise matrix, and 
%In this paper, integrated NF localization and communication is formulated as a matrix factorization problem with proper constraints on the factor matrices, i.e., one factor matrix is sparse and the columns of the other factor matrix are subject to the structure, specified by the distance-angle dependent steering vectors. 
To solve a generic MF problem, \rew{leveraging varitional inference (VI) \cite{winn2005variational} and unitary approximate message passing (UAMP) \cite{guo2015approximate, yuan2021BiUTAMP}, we develop an efficient Bayesian approach. In choosing the variational distribution, instead of using the mean field approximation with full factorization, we only decouple $\bA$ and $\bX$ and treat them as two latent matrices to avoid performance loss, leading to the updates of two distributions on matrices $\bA$ and $\bX$, which are difficult and expensive.} %due to the various priors and matrix forms of $\bX$ and $\bH$. 
%With matrix \rew{Gaussian} distributions \cite{Waal1985} and
\rew{By exploiting the structure of the variational messages on $\bA$ and $\bX$  and} through a covariance matrix whitening process, we incorporate UAMP into VI to efficiently deal with the updates of the distributions of $\bA$ and $\bX$. The VI-based method is implemented using message passing with UAMP as its key component, leading to an algorithm called UAMP-MF.  
%leveraging UAMP, we propose a new Bayesian algorithm called UAMP-MF for MF. The algorithm is developed based on variational inference (VI) \cite{winn2005variational} (in a matrix form derived in this paper) and color noise whitening, where matrix normal distribution \cite{Waal1985} %\cite{Glanz2013} is used and UAMP serves as the key component of the UAMP-MF algorithm to achieve high efficiency and robustness.
UAMP-MF inherits the low complexity and robustness of UAMP. %It has cubic complexity per iteration, which is considered low for an MF problem (note that even the multiplication of two matrices has cubic complexity).
\rew{Enjoying the flexibility of a Bayesian approach and low complexity and robustness of UAMP, UAMP-MF can handle various MF problems in a unified way while with high computational efficiency.} We then apply the developed UAMP-MF algorithm to solve the JNFLSD problem, where the structures of the factor matrices are fully exploited. Extensive simulation results are provided to demonstrate the superior performance of the proposed method.

The remainder of this paper is organized as follows. In Section II, we introduce the signal model for integrated near-field localization and communications, and formulate JNFLSD as a matrix factorization problem. In Section III, we develop the matrix factorization algorithm UAMP-MF, leveraging VI and UAMP. In Section IV, the UAMP-MF algorithm is applied to tackle the JNFLSD problem. Simulation results are provided to demonstrate the superiority of the UAMP-MF based algorithm in Section V, and conclusions are drawn in Section VI.

{Throughout the paper, we use the following notations. Boldface lower-case and upper-case letters denote vectors and matrices, respectively. %and superscript $(\cdot)^T$ represents the transpose operation.
	A Gaussian distribution of $x$ with mean {$\hat x$} and variance $\nu_x$ is represented by $\N(x;{\hat x},\nu_x)$.
	%We also simply use $\N(m, v)$ to represent a Gaussian distribution with mean $m$ and variance $v$.
	Notation
	%$\otimes$ represents the Kronecker product, and
	$\Tr(\cdot)$ denotes the trace operation.
	The relation $f(x)=cg(x)$ for some positive constant $c$ is written as $f(x)\propto g(x)$, and $diag(\ba)$ returns a diagonal matrix with $\ba$ on its diagonal. We use $\bA\cdot\bB$ and $\bA\cdot/\bB$ to denote the element-wise product and division between $\bA$ and $\bB$, respectively. %The notation $\ba^{.-1}$ denotes the element-wise inverse operation to vector $\ba$.
	The notation $|\bA|^{.2}$ denotes element-wise magnitude squared operation for $\bA$, and $||\bA||$ is the Frobenius norm of $\bA$. %The notation $<\ba>$ denotes the average operation for $\ba$, i.e., the sum of the elements of $\ba$ divided by the number of its elements.
	%The notation $\int_{\bc\vee {c_n}} f_{\bc}(\bc)$ represents integral over all elements in $\bc$ except $c_n$.
	We use $\textbf{1}$, $\textbf{0}$ and $\bI$ to denote an all-one matrix, an all-zero matrix and an identity matrix with a proper size, respectively. The above operations defined for matrices are applied to vectors. We use $\tVec(\cdot)$ to denote the vectorization operation.
We use $\MN\big(\bX;\hat\bX,\bU_X,\bV_X\big)$ to denote the matrix \rew{Gaussian} distribution, which is a generalization of the multivariate \rew{Gaussian} distribution to matrix-valued random variables \cite{Waal1985},
where \rew{$\bX$ is a random Gaussian matrix}, $\hat\bX$ is the mean of $\bX$,  $\bU_X$ and $\bV_X$ are the covariance among rows and columns of $\bX$, respectively. The matrix \rew{Gaussian} distribution is related to the multivariate \rew{Gaussian} distribution in the way that  $\bx \sim \N\big(\bx;\hat\bx,\bV_X\otimes \bU_X\big)$, where $\bx=\tVec(\bX)$ and $\hat\bx=\tVec(\hat\bX)$.

\section{Signal Model and Problem Formulation for Integrated Near Field Sensing and Communications}

%\subsection{Signal Model and Problem Formulation}

%In this section, we apply the UAMP-MF algorithm to address the integrated sensing and communication problem, and develop an efficient message passing algorithm for user joint user localization and signal detection.

We consider an uplink near-field ISAC system, where a BS equipped with $R$ antennas provides communications and localization services to $K$ single antenna users. In particular, we focus on uplink transmission, where users transmit signals to the BS, and the BS performs JNFLSD based on the received signals. It is worth highlighting that the method developed in this paper is automatically applicable to the far-field scenario because the near-field signal model degenerates to the far-field one when the distance from the BS to users is large. In addition, we assume that differentiation modulation is employed by all users, enabling that JNFLSD is realized without using pilot signals. This is significant as the training period can be skipped and considerable pilot overhead can be saved, thereby making the system attractive for the scenario of moving mobile terminals.    

\begin{figure}[!t]
	\centering
	\includegraphics[width=0.4\textwidth]{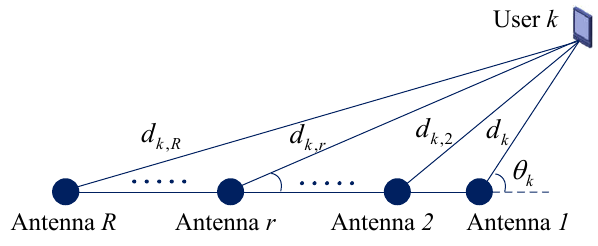}
	\centering
	\caption{Illustration of near-field angle-distance dependent signal model (only a single user is shown).}
	\label{fig:NearFieldModel}
\end{figure}

As shown in Fig. \ref{fig:NearFieldModel}, for simplicity, we assume that a uniform linear array is used by the BS. However, we note that the method proposed in this paper can be readily extended to other array configurations. We assume that the antenna spacing is $\lambda/2$, resulting in an aperture of $D = (R-1)\lambda/2$, where $\lambda$ is the signal wavelength. The boundary between near-field and far-field is determined by the Rayleigh distance $2D^2/\lambda$ \cite{Selvan2017}. Because of the use of antenna arrays with large aperture at the BS and the short wavelength of mmWave/Terahertz band signals, there is a high probability that the users locate in the near-field of the BS. Due to the high attenuation of mmWave/Terahertz band signals, we only consider the line of sight (LOS) path between a user and the BS \cite{Rappaport2017}. The distance between the $k$th user and the first antenna (reference point) of the antenna array at the BS is denoted by $d_k$, and $\theta_k$ represents the angle between the antenna array and the $k$th user, as shown in Fig. \ref{fig:NearFieldModel}. The received signal at the $r$th antenna element from the $k$th user at time instant $l$ can be expressed as
\begin{eqnarray}
	y_{r,l}&=&\exp\left(\frac{-j2\pi}{\lambda}(d_{k,r}-d_{k})\right) x_{k,l} + w_{r,l}
	%\nonumber \\ &=&\alpha_r(d_k,\theta_k)x_{kl} + w_{r,l},
\end{eqnarray}
where $d_{k,r}$ denotes the distance between the $k$th user to the $r$th antenna at the BS, {$x_{k,l}=g_ks_{k,l}$ with $g_k$ being the channel gain of user $k$ (which is a constant over a short time period) and $s_{k,l}$ being the transmitted signal of user $k$, and $w_{r,l}$ denotes the noise.} %$\que{\alpha_r(d_k,\theta_k)= .....}$ will be used to form an angle-distance-dependent steering vector.
According to Fig. \ref{fig:NearFieldModel}, %and $\alpha_r(d_k,\theta_k)$ can be expressed as
\begin{equation}
	d_{k,r}-d_{k}=\sqrt{d_k^2+b_r^2+2d_k b_r \cos\theta_k}-d_k
\end{equation}
with $b_r=(r-1)\lambda/2$, which is related to the angle $\theta_k$. Hence we define
\begin{eqnarray}
	a_r(d_k,\theta_k) \!\!\! \!\!\!&=& \!\!\!\!\!\! \exp\left(\frac{-j2\pi}{\lambda}(d_{k,r}-d_{k})\right) \nonumber \\
	\!\!\!\!\!\!&=&\!\!\!\!\!\!\exp\!\left(\!\frac{-j2\pi}{\lambda}\!\!\!\left[\sqrt{d_k^2+\!b_r^2+\! 2d_k b_r \cos\theta_k}-d_k\right]\right).
\end{eqnarray}
Then, for all the antenna elements, the angle-distance dependent steering vector can be represented as
\begin{equation}
	\ba(d_k,\theta_k)=[a_1(d_k,\theta_k),...,a_R(d_k,\theta_k)]\tra.
\end{equation}

Consider $K$ users in the system, and the received signal by the BS for all users in a time duration of $L$ sampling time instants can be expressed as
\begin{eqnarray}
	\bY=\bA\bX+\bW, \label{eq:recv1}
\end{eqnarray}
where $\bX=\left[\bx_1,...,\bx_K\right]\tra\in\mathbb{C}^{K\times L}$, $\bW\in\mathbb{C}^{R\times L}$ is the white Gaussian noise matrix, and
\begin{equation}
	\bA=\left[\ba(d_1,\theta_1),...,\ba(d_K,\theta_K)\right]\in \mathbb{C}^{R\times K}. \label{eq:ConstructA}
\end{equation}

%\subsection{Problem Formulation}

The aim of the BS is to locate the users by estimating the parameters $\{d_k,\theta_k\}$ and detect the transmitted symbols of all users based on the estimates of $\{x_{k,l}\}$. As no pilot signals are used, only the received signal matrix $\bY$ is available at the BS. We have the following remarks: 
\begin{itemize}
\item We need to factorize the received signal matrix $\bY$ to the product of matrices $\bA$ and $\bX$, plus noise matrix $\bW$, thereby the estimates of the matrices $\hat \bA$ and  $\hat \bX$ can be obtained. This is an MF problem.
\item It is noted that the columns of matrix $\bA$ are parameterized by parameters $\{d_k,\theta_k\}$, the estimation of matrix $\bA$ produces the parameter estimates. As differential modulation is employed, the transmitted symbols of the users can be detected based on the estimate $\hat \bX$.
\item  A conventional way to solve the problem is transforming the problem to a sparse signal recovery one by constructing a dictionary matrix $\bA'$ with a two-dimensional grid over $\{d_k,\theta_k\}$, leading to a sparse matrix $\bX'$ with model $\bY=\bA'\bX'+\bW$. With the known dictionary matrix $\bA'$, one only needs to recovery the sparse matrix $\bX'$. However, this way is normally impractical for the considered problem because the size of the dictionary matrix is huge as the grid is established over two parameters $d_k$ and $\theta_k$, resulting in prohibitively high complexity. In addition, the true $\{d_k,\theta_k\}$ may not land on the grids, rendering grid mismatch errors. Off-grid methods may be used to deal with the grid-mismatch problem, but they still require the assist of a grid, which is huge for the considered problem. So the complexity is a serious concern. 
\end{itemize}	

In this paper, we address the problem from the perspective of MF, where the matrices $\bA$ and $\bX$ are estimated simultaneously. We start from the development of a highly efficient algorithm UAMP-MF to solve a generic MF problem, and then apply it to tackle the blind JNFLSD problem.

\section{Matrix Factorization Using Unitary Approximate Message Passing}
%with the following model
%\begin{align}
%	\bY=\bA\bX+\bW,  \label{eq:bilinear}
%\end{align}
%where $\bY\in\mC^{M\times L}$ denotes a known matrix, $\bW\in\mC^{M\times L}$ accounts for unknown perturbations or measurement errors,
%that is disturbed by an additive white Gaussian noise matrix $\bW$,
%and $\bA\in\mC^{M\times N}$ and $\bX\in\mC^{N\times L}$ are two factor matrices to be obtained.

We consider a generic MF problem with model \eqref{eq:recv1}, and abuse the use of the notation $\bA$ and $\bX$ for more general scenarios, i.e., depending on concrete application scenarios, the matrices $\bA$ and $\bX$ are subject to some structures. For instance, in dictionary learning (DL) \cite{Rubinstein2010DL}, matrix $\bX$ is sparse matrix and $\bA$ is a dictionary matrix to be learned. In compressive sensing with matrix uncertainty (CSMU) \cite{Zhu2011CSMU}, $\bX$ is sparse and the sensing matrix $\bA$ can be modeled as $\bA=\bar \bA +\delta\bA$, where the matrix $\bar \bA$ is known, and $\delta\bA$ denotes an unknown perturbation matrix. The robust principal component analysis (RPCA) problem \cite{Cand2011} can also be formulated as \eqref{eq:recv1}, where both $\bA$ and $\bX$ admit specific structures \cite{Parker2014I}. We may also be interested in sparse MF, where both $\bA$ and $\bX$ are sparse. In this section, we first give a brief introduction to the VI and (U)AMP, then develop the UAMP-MF algorithm for a generic MF problem by incorporating UAMP to VI. The use of the UAMP-MF algorithm to solve the formulated JNFLSD problem will be elaborated in Section IV. 
%leveraged to develop the matrix factorization algorithm UAMP-MF. The UAMP-MF algorithm forms the basis to solve the integrated near field sensing and communications problem in Section III.

\subsection{Variational Inference}

VI is a machine learning method widely used to approximate posterior densities for Bayesian models \cite{VIreview}, \cite{winn2005variational}, \cite{VImessagepassing}. Let $\bV$ and $\bR$ be the set of hidden (latent) variables and visible (observed) variables, respectively, with joint distribution $p(\bV,\bR)$. The goal of VI is to find a tractable variational distribution ${q}(\bV)$ that approximates the true posterior distribution $p(\bV|\bR)$. With the distribution $q(\bV)$, the log marginal distribution of the observed variables admits the following decomposition
\begin{equation}
	\ln p(\bR)=\mathcal{L}(q(\bV)) + \mathcal{KL}(q(\bV)||p(\bV|\bR)),
\end{equation}
where the variational lower bound $\mathcal{L}(q(\bV))$ is given as
\begin{equation}
	\mathcal{L}(q(\bV))=\int_{\bV} q(\bV) \ln \frac{p(\bV, \bR)}{q(\bV)},
\end{equation}
and the Kullback-Leibler (KL) divergence between $q(\bV)$ and $p(\bV|\bR)$ is
\begin{equation}
	\mathcal{KL}(q(\bV)||p(\bV|\bR))=-\int_{\bV} q(\bV) \log \frac{p(\bV| \bR)}{q(\bV)}.
\end{equation}
The distribution $q(\bV)$ that minimizes the KL divergence $\mathcal{KL}(q(\bV)||p(\bV|\bR))$ can be found by maximizing the variational lower bound $\mathcal{L}(q(\bV))$.

VI can be implemented \rew{using} message passing with the assistance of graphical models \rew{\cite{winn2005variational}, \cite{VImessagepassing}}. If the variational distribution with some factorization is chosen, e.g.,
\begin{equation}
	q(\bV)=\prod_k q_k(\bV_k),
\end{equation}
where $\bV=\{\bV_k\}$,
then the variational distribution can be found through an iterative procedure \cite{winn2005variational} with the update rule
\begin{equation}
	q_k(\bV_k) \propto \exp \left(\int_{ \tilde {\bV}} q(\tilde{\bV}) \log f(\bV_k, \tilde{\bV})  \right).
\end{equation}
Here $ f(\bV_k, \tilde{\bV})$ is a local factor associated with $\bV_k$ and $\tilde{\bV} \in \{\bV_i, i \neq k \} $,
%may only include some of the variables in $\{\mathbf{V}_i, i \neq k \}$,
depending on the structure of the factor graph. The updates of $\{q_k(\bV_k)\}$ are carried out iteratively until it converges or a pre-set number of iterations is reached.

\subsection{(U)AMP}

AMP was derived based on loopy BP with Gaussian and Taylor-series approximations \cite{donoho2009message}, {\cite{rangan2011generalized}}, which can be used to recover $\bx$ from the noisy measurement $\by=\bA\bx+\bw$ with $\bw$ being a zero-mean white Gaussian noise vector. It works well for an i.i.d. (sub)Gaussian $\bA$, but can easily diverge for generic $\mathbf{A}$ \cite{rangan2019convergence}.
%Inspired by \cite{guo2013}, the work
\rew{It is shown in \cite{guo2015approximate} that the robustness of AMP can be significantly improved through simple pre-processing, i.e., performing a unitary transformation to the original linear model  \cite{yuan2021BiUTAMP}.
With an SVD  $\bA= \bU \boldsymbol{\Lambda} \bV$,
%where $\bU$ and $\bV$ being two unitary matrices,
performing a unitary transformation with $\bU^H$ leads to the following model}
\begin{equation}
\br=\mathbf{\Phi} \bx+\boldsymbol{\omega},
\label{r=uy}
\end{equation}
where $\br=\bU^H \by$, $\boldsymbol{\Phi}=\bU^H\bA=\boldsymbol{\Lambda}\bV$,
$\mathbf{\Lambda}$ is %an $M\times N$
rectangular diagonal matrix, and $\boldsymbol{\omega} = \bU^H \bw$ remains \rew{white and Gaussian.} %which has the same mean and covariance as $\boldsymbol{\omega}$.} %as $\mathbf{U}^H$ is a unitary matrix.

Applying the vector step size AMP {\cite{rangan2011generalized}} with model \eqref{r=uy} leads to the first version of UAMP {(called UAMPv1)} shown in Algorithm \ref{UTAMPv1} \footnote{\rew{Replacing $\br$ and $\mathbf{\Phi}$ with $\by$ and $\bA$, the original AMP algorithm is recovered.}}. An average operation can be applied to two vectors: $\boldsymbol{\tau}_x$ in Line 7 and $|\mathbf{\Phi}^H |^{.2}  \boldsymbol{\tau}_s$ in Line 5 of {UAMPv1 in} Algorithm \ref{UTAMPv1}, leading to the second version of UAMP \cite{guo2015approximate} {(called UAMPv2), where the operations in the brackets of Lines 1, 5 and 7 are executed} (refer to \cite{yuan2021BiUTAMP} for details).
%Compared to AMP and UAMPv1,  UAMPv2 does not require matrix-vector multiplications in Lines 1 and 5, so that the number of matrix-vector products is reduced from 4 to 2 per iteration. %This is a significant reduction in computational complexity as the complexity of AMP-like algorithms is dominated by matrix-vector multiplications.

\begin{algorithm}
	\caption{UAMP (UAMPv2 executes operations in [ ])}
	%	Unitary transform: $\mathbf{r=U}^H \mathbf{ y }=\mathbf{\Phi x} +\boldsymbol{\omega}$, where $\mathbf{\Phi}=\mathbf{U}^H\mathbf{A}=\mathbf{\Lambda V}$, and $\mathbf{U}$ is obtained from the SVD $\mathbf{A=U \Lambda V}$.\\
	Initialize $\boldsymbol{\tau}_x^{(0)} (\mathrm{or}~{\tau}_x^{(0)})>0$ and ${{\bx}^{(0)} }$. Set $\bs^{(-1)}=\mathbf{ 0 }$ and $t=0$. Define vector $\boldsymbol{\lambda}={\mathbf{ \Lambda \Lambda}^H} \textbf{1}$.\\
	\textbf{Repeat}
	\begin{algorithmic}[1]
		\STATE $\boldsymbol{\tau}_p$ = $   \mathbf{|\Phi|}^{.2} \boldsymbol{\tau}^t_x$~~~~~~~~~~~~ $\left[\mathrm{or}~ \boldsymbol{\tau}_p =  \tau^t_x  \boldsymbol{\lambda}\right]$ \\
		\STATE $ \bp= \mathbf{\Phi}  {{\bx}^{t} } - \boldsymbol{\tau}_{p} \cdot  \bs^{t-1} $\\
		\STATE $ \boldsymbol{\tau}_s = \mathbf{1}./ (\boldsymbol{\tau}_p+\beta^{-1} \mathbf{1}) $\\
		\STATE $ \bs^t= \boldsymbol{\tau}_s \cdot (\br-\bp) $\\
		\STATE $\mathbf{1}./ \boldsymbol{\tau}_q$ = $ |\mathbf{\Phi}^H |^{.2}  \boldsymbol{\tau}_s$~~~~~~~$\left[\mathrm{or}~ \mathbf{1}./ \boldsymbol{\tau}_q = (\frac{1}{N} \boldsymbol{\lambda}^H \boldsymbol{\tau }_s) \mathbf{1}\right]$ \\
		\STATE $ \bq = {{\bx}^{t} } + \boldsymbol{\tau}_q \cdot(  \mathbf{\Phi}^H \bs^t)$\\
		\STATE $\boldsymbol{\tau}_x^{t+1}$ = $\boldsymbol{\tau}_q \cdot g_{x}' ( \bq, \boldsymbol{\tau}_q)$~~~~~~$\left[\mathrm{or}~ \tau_x^{t+1} \!=\!  \frac{1}{N}  \mathbf{1}^H  \left(\boldsymbol{\tau}_q \cdot g_{x}' ( \bq, \tau_q)\right) \right]$ \\
		\STATE 	$ {{\mathbf{x}}^{t+1} } = g_{x}  ( \bq, \boldsymbol{\tau}_q)$\\	
		\STATE 	$  t=t+1$
	\end{algorithmic}
	\textbf{Until terminated}
	\label{UTAMPv1}
\end{algorithm}

%Interestingly, the average operations can also enhance the stability of the algorithm. UAMP version 2 converges for any matrix $\mathbf{A}$ in the case of Gaussian priors \cite{guo2015approximate}.

In the (U)AMP algorithm, $g_x(\bq, \boldsymbol{\tau}_q )$ is related to the prior of $\bx$, \rew{which is a column vector with the} $n$th
entry %$[ g_x(\mathbf{q}, \boldsymbol{\tau}_q ) ]_n$
given by
\begin{equation}
[g_x(\bq, \boldsymbol{\tau}_q ) ]_n
=
\frac{\int x_n p(x_n) \mathcal{N} (x_n ; q_n, \tau_{q_n})  d x_n }{\int  p(x_n) \mathcal{N} (x_n ; q_n, \tau_{q_n})  d x_n },
\label{g_x}
\end{equation}
where  $p(x_n)$ represents a prior for $x_n$, \rew{and $q_n$ and $\tau_{q_n}$ are the $n$th entry of $\bq$ and $\boldsymbol{\tau}_q$, respectively.}
The function $g_x'(\bq,\boldsymbol{\tau}_q)$ returns a column vector and the $n$th element is denoted by $[ g_x'(\bq, \boldsymbol{\tau}_q ) ]_n$, where the derivative is taken with respect to $q_n$.

\subsection{Design of UAMP-MF}

\rew{With the Bayesian treatment of MF by transforming the constraints on matrices $\bA$ and $\bX$ to their priors $p(\bA)$ and $p(\bX)$ properly, many MF problems can be handled in a unified way. In this work, we assume that the priors are separable, i.e., $p(\bA)=\prod\nolimits_{m,n}p(h_{m,n})$ and $p(\bX)=\prod\nolimits_{n,l}p(x_{n,l})$, which can be used for NMF, DL, CSMU, RPCA and sparse MF, as shown in Section IV.}

\rew{%Besides $\bA$ and $\bX$, we also assume that the noise precision is unknown.
With model \eqref{eq:recv1}, we have the following joint conditional distribution and its factorization}
\begin{eqnarray}
	p(\bX,\bA,\lambda|\bY) \!\!\!\!\!&\propto&\!\!\!\!\!  p(\bY|\bX,\bA,\lambda)p(\bX)p(\bA)p(\lambda), \label{eq:factor}
	%\!\!\!\!\!&\triangleq&\!\!\!\!\! f_{Y} (\bY, \! \bX, \!\bA, \!\lambda) f_{X}(\bX) f_{H} (\bA) f_{\lambda} (\lambda)
\end{eqnarray}
\rew{where we assume that the entries of $\bW$ are i.i.d. Gaussian with zero mean precision $\lambda$ (this is applied to all the examples in this paper), and}
\begin{eqnarray}
%	f_{Y} (\bY, \bX, \bA, \lambda) %&\triangleq&
p\big(\bY|\bX,\bA,\lambda\big)
	= \MN\big(\bY;\bA\bX,\bI_M,\lambda^{-1}\bI_L\big).
\end{eqnarray}
We assume a Jefferys prior
%$f_{\lambda}(\lambda)\triangleq p(\lambda)\propto 1/\lambda$
$p(\lambda)\propto 1/\lambda$
\cite{Tipping} for the noise precision. %and the priors for $\bA$ and $\bX$ denoted by $f_{H}(\bA)\triangleq p(\bA)$ and $f_{X}(\bX)\triangleq p(\bX)$ are customized according to the requirements in a specific problem.

\rew{If the a posteriori distributions $p(\bA|\bY)$ and $p(\bX|\bY)$ can be found, then the estimates of $\bA$ and $\bX$ can be obtained, e.g., using the a posteriori means of $\bA$ and $\bX$ to serve as their estimates. However, it is intractable to find the exact a posteriori distributions in general, so we resort to VI to find their variational approximations. We define a variational distribution}
\begin{equation}
	q(\bX, \bA, \lambda)=q(\bX)q(\bA)q(\lambda),
\end{equation}
\rew{and expect that $q(\bX) \approx p(\bX|\bY)$, $q(\bA)\approx p(\bA|\bY)$ and $q(\lambda) \approx p(\lambda|\bY)$. However, finding the varational distributions is still challenging due to the high dimensions of $\bA$ and $\bX$, and the priors of $\bA$ and $\bX$ may also lead to intractable $q(\bA)$ and $q(\bX)$.}
\rew{In this work, %through a covariance matrix whitening process,
UAMP is employed to solve these challenges with high efficiency. This also leads to Gaussian approximations to $q(\bA)$ and $q(\bX)$, so that the estimates of $\bA$ and $\bX$ (i.e., the a posteriori means of $\bA$ and $\bX$) appear as the parameters of the distributions.
%The factor graph representation of $\eqref{eq:factor}$ is depicted in Fig.\ref{fig:FactorGraph}.
Leveraging UAMP, we carry out VI in a message passing manner, with the aid of a factor graph representation of the problem, depicted in Fig.\ref{fig:FactorGraph}. This leads to the message passing algorithm UAMP-MF.}

\rew{Next, we derive the UAMP-MF algorithm and show how to efficiently update $q(\bX)$, $q(\bA)$ and $q(\lambda)$ iteratively. In the derivation of UAMP-MF, we use the notation $m_{n_a \rightarrow  n_b} (\bX)$ to denote a message passed from node $n_a$ to node $n_b$, which is a function of $\bX$. The  UAMP-MF algorithm is shown in Algorithm \ref{alg:UAMPMF}, which we will frequently refer to in this section. }

\begin{figure}[!t]
	\centering
	\includegraphics[width=0.30\textwidth]{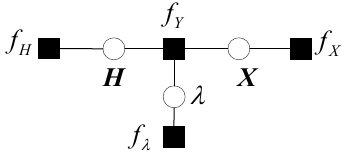}
	\centering
	\caption{Factor graph of $\eqref{eq:factor}$, \rew{where $f_A \triangleq p(\bA)$, $f_{Y} \triangleq p(\bY|\bX,\bA,\lambda)$, $f_X \triangleq p(\bX)$, and $f_{\lambda}\triangleq p(\lambda)$.}}
	\label{fig:FactorGraph}
\end{figure}

\begin{algorithm}
	\caption{UAMP-MF}
	\textbf{Initialization}: $\bU_A=\bI$, $\bV_A=\bI$, $\hat\bA=\textbf{1}$, $\bV_X=\bI$.
	$\bXi_X= \textbf{1}$, $\bS_X=\textbf{0}$,
	$\bXi_A=\textbf{1}$, and $\bS_A=\textbf{0}$. \\
	\textbf{Repeat}
	\begin{algorithmic}[1]
		\STATE$\overline\bW_X=\hat\bA\herm\hat\bA+ M \bV_A$ \label{alg:UAMPMF_X1}
		\STATE$[\bC_X,\bD_X]=\text{eig}(\overline\bW_X)$
		\STATE$\bR_X = \bD_X^{-\frac{1}{2}}\bC_X\herm\hat\bA\herm\bY$,
		$\bPhi_X= \bD_X^{-\frac{1}{2}}\bC_X\herm$
		\STATE$\bV_{P_X}=|\bPhi_X|^{.2}\bXi_X$
		\STATE $\bP_X=\bPhi_X\hat\bX-\bV_{\bP_X}\cdot\bS_X$
		\STATE$\bV_{S_X}=\textbf{1}./(\bV_{P_X}+\hat\lambda^{-1}\mathbf{1})$
		\STATE$\bS_X=\bV_{S_X}\cdot(\bR_X-\bP_X)$
		\STATE$\bV_{Q_X}=\textbf{1}./(|\bPhi_X\herm|^{.2}\bV_{S_X})$
		\STATE$\bQ_X=\hat\bX+\bV_{Q_X}\cdot(\bPhi_X\herm\bS_X)$ \label{alg:UAMPMF_Qx}
		\STATE{$\bXi_X=\bV_{Q_X}\cdot\bG_X'(\bQ_X,\bV_{Q_X})$}  \label{alg:UAMPMF_VQx}
		\STATE$\hat\bX=\bG_X(\bQ_X,\bV_{Q_X})$   \label{alg:UAMPMF_Hatx}
		\STATE$\bU_X=\text{diag}(\text{mean}(\bXi_X,2))$ \label{alg:UAMPMF_Ux}
		\STATE$\overline\bW_A=\hat\bX\hat\bX\herm+ L\bU_X$ \label{alg:UAMPMF:H1}
        \STATE$[\bC_A,\bD_A]=\text{eig}(\overline\bW_A)$
		\STATE$\bR_A = \bD_A^{-\frac{1}{2}}\bC_A\herm\hat\bX\bY\herm$,
		$\bPhi_A = \bD_A^{-\frac{1}{2}}\bC_A\herm$
		\STATE$\bV_{P_A}=|\bPhi_A|^{.2}\bXi_A\herm$
		\STATE$\bP_A=\bPhi_A\hat\bA\herm-\bV_{\bP_A}\cdot\bS_A$
		\STATE$\bV_{S_A}=\mathbf{1}./(\bV_{P_A}+\hat\lambda^{-1}\mathbf{1})$
		\STATE$\bS_A=\bV_{S_A}\cdot(\bR_A-\bP_A)$
		\STATE$\bV_{Q_A}=\mathbf{1}./(|\bPhi_A\herm|^{.2}\bV_{S_A})$ \label{alg:UAMPMF_VQA}
		\STATE$\bQ_A=\hat\bA\herm+\bV_{Q_A}\cdot(\bPhi_A\herm\bS_A)$ \label{alg:UAMPMF_QA}
		\STATE{$\bXi_A=\bV_{Q_A}\cdot\bG_A'(\bQ_A\herm,\bV_{Q_A}\tra)$}
		\STATE $\hat\bA=\bG_A(\bQ_A\herm,\bV_{Q_A}\tra)$
		\STATE$\bV_A=\text{diag}(\text{mean}(\bXi_A,1))$ \label{alg:UAMPMF_VH}
		\STATE$\hat\lambda=ML/C$ with $C$ given in \label{alg:UAMPMF_C}
	\end{algorithmic}
	\textbf{Until terminated}
	\label{alg:UAMPMF}
\end{algorithm}

\subsubsection{Update of $q(\bX)$}
According to VI, with $q(\bA)$ and $q(\lambda)$ (updated in the last iteration), we compute $q(\bX)$. As shown by the factor graph in Fig. \ref{fig:FactorGraph}, we need to compute the message $m_{f_Y\to X}(\bX)$ from the factor node $f_Y$ to the variable node $\bX$ and then combine it with the prior $p(\bX)$. Later we will see that $q(\bA)$ is a matrix \rew{Gaussian distribution}, i.e., $q(\bA)=\MN(\bA;\hat\bA,\bU_A,\bV_A)$ with \rew{a mean matrix $\hat\bA$ (see Line 23 of Algorithm 2), a column covariance matrix $\bV_A$ and row covariance matrix $\bU_A=\bI_{N}$}, and $q(\lambda)$ is a Gamma distribution. It turns out that  $m_{f_Y\to X}(\bX)$ is matrix \rew{Gaussian}, shown by Proposition 1.

\vspace{0.1 in}
\textbf{Proposition 1}: The message from $f_Y$ to $\bX$ can be expressed as a matrix \rew{Gaussian} distribution, i.e., %of $\bX$, which can be expressed as
\begin{equation}
m_{f_Y\to X}(\bX)\propto \MN(\bX;\overline\bX, \hat\lambda^{-1}\overline\bU_X,\bI_L), \label{eq:msg_fyX_result}
\end{equation}
with
\begin{equation}
\overline\bX = \overline\bU_X\hat\bA\herm\bY, \label{eq:msg_fyX_mean}
\end{equation}
\begin{equation}
\overline\bU_X = \big(\hat\bA\herm\hat\bA+\Tr(\bU_A)\bV_A\big)^{-1}, \label{eq:msg_fyX_U}
\end{equation}
and
\begin{equation}
\hat\lambda=\int_{\lambda} \lambda q(\lambda).
\end{equation}
The computation of $\hat \lambda$ is shown in \eqref{eq:lambda}.
 %is the mean of $\lambda$ based on the distributions $q(\lambda)$, i.e., .

%and $\hat\bA$ are respectively the

\begin{proof}
See Appendix A.
\end{proof}

%In the above, we have obtained the message $m_{f_Y\to X}(\bX)$.
\rew{Then the message  $m_{f_Y\to X}(\bX)$ needs to be combined with the prior $p(\bX)$ to obtain $q(\bX)$. This can be challenging as $\bX$ is a random matrix with high dimension, and the prior $p(\bX)$ may lead to an intractable $q(\bX)$ and high computational complexity. Next, leveraging UAMP, we update $q(\bX)$ efficiently.} In addition, we will also circumvent the matrix inversion involved in \eqref{eq:msg_fyX_U}.

Note that $\bX=[\bx_1,...,\bx_l]$ and $\overline\bX=[\overline\bx_1,...,\overline\bx_l]$. The result in \eqref{eq:msg_fyX_result} indicates that $\bx_l \sim \N(\bx_l;\overline\bx_l, \hat\lambda^{-1}\overline\bU_X)$, \rew{and all vectors in $\bX$ have a common covariance matrix, which will greatly simplify the computations later. With the result, for each $\bx_l$, we have} the following pseudo observation model
\begin{eqnarray}
\overline\bx_l=\bx_l+\be_l, \label{eq:pseudoX1}
\end{eqnarray}
where $\be_l\sim \N(\be_l; 0,\hat\lambda^{-1}\overline\bU_X)$, i.e., the model noise is not white. We can whiten the noise by left-multiplying  %\triangleq (\vec\bU_X^{-1})^{\frac{1}{2}}$, then
both sides of \eqref{eq:pseudoX1} by $\overline\bU_X^{-\frac{1}{2}}$, leading to
\begin{eqnarray}
\overline\bU_X^{-\frac{1}{2}}\overline\bx_l
%&=&\overline\bU_X^{-\frac{1}{2}}\bx_l+\overline\bU^{-\frac{1}{2}}\be_x\nonumber\\
= \overline\bU_X^{-\frac{1}{2}}\bx_l+\bw_l,
\end{eqnarray}
where $\bw_l=\overline\bU_X^{-\frac{1}{2}}\be_l$ is white and Gaussian with covariance $\hat\lambda^{-1}\bI$.  Through the whitening operation, we get a standard linear model with white additive Gaussian noise, which facilitates the use of UAMP. Considering all the vectors in $\overline\bX$ \rew{which share the same whitening matrix}, we have  %This processing can be named as whitening step. So \eqref{eq:msg_fyX_U} can be rewritten as matrix form
\begin{eqnarray}
\overline\bU_X^{-\frac{1}{2}}\overline\bX=\overline\bU_X^{-\frac{1}{2}}\bX+ \boldsymbol{\Omega_X}, \label{eq:pseudoX2}
\end{eqnarray}
where $\boldsymbol{\Omega_X}$ is white and Gaussian.

With model \eqref{eq:pseudoX2} and the prior $p(\bX)$, we use UAMP to update $q(\bX)$. Following UAMP, a unitary transformation needs to be performed with the unitary matrix $\bC_X\herm$ obtained from the SVD of $\overline\bU_X^{-\frac{1}{2}}$ (or eigenvalue decomposition (EVD) as the matrix is definite and symmetric), i.e.,
\begin{eqnarray}
\overline\bU_X^{-\frac{1}{2}} = \bC_X \bLambda\bC_X\herm
\end{eqnarray}
with $\bLambda$ being a diagonal matrix. After performing the unitary transformation, we have
\begin{eqnarray}
\bR_X=\bPhi_X \bX+ \boldsymbol{\Omega'_X},\label{eq:pseudoX3}
\end{eqnarray}
where $\rew{\bR_X}=\bC_X\herm\overline\bU_X^{-\frac{1}{2}}\overline\bX$, $\bPhi_X=\bC_X\herm\overline\bU_X^{-\frac{1}{2}}=\bLambda\bC_X\herm$  and $ \boldsymbol{\Omega'_X}=\bC_X\herm \boldsymbol{\Omega_X}$, which is still Gaussian and white.

From the above, \rew{the direct way to obtain $\bR_X$ and $\bPhi_X$ in model \eqref{eq:pseudoX3} are costly}. Specifically, a matrix inverse operation needs to be performed according to \eqref{eq:msg_fyX_U} so that $\overline \bX$ in \eqref{eq:msg_fyX_mean} can be obtained; a matrix squared root operation is needed to obtain $\overline\bU_X^{-\frac{1}{2}}$; and an SVD operation is required to $\bC_X\herm$, \rew{which lead to high complexity. Next, we show that these expensive computations can be avoided with an EVD operation.}

Instead of computing $\overline\bU_X^{-1}$ and $\overline\bU_X^{-\frac{1}{2}}$ followed by SVD of $\overline\bU_X^{-\frac{1}{2}}$, we perform EVD to $\overline\bU_X^{-1}=\overline\bW_X=\hat\bA\herm\hat\bA+\Tr(\bU_A)\bV_A$, i.e.,
\begin{eqnarray}
	[\bC_X,\bD_X]=\text{eig}(\overline\bW_X),
\end{eqnarray}
where the diagonal matrix
\begin{equation}
	\bD_X= \bLambda^{-2}.
\end{equation}
Hence $\bPhi_X=\bLambda\bC_X\herm=\bD_X^{-1/2}\bC_X\herm$, where the computation of $\bD_X^{-1/2}$ is trivial as $\bD_X$ is a diagonal matrix. Meanwhile, the computation of the pseudo observation matrix $\bR_{X}$ can also be simplified:
\begin{eqnarray}
\bR_X&=&\bC_X\herm\overline\bU_X^{-\frac{1}{2}}\overline\bX \nonumber \\
&=&\bC_X\herm\overline\bU_X^{-\frac{1}{2}}\overline\bU_X\hat\bA\herm\bY \nonumber\\
&=& {\bD_X^{-\frac{1}{2}}\bC_X\herm\hat\bA\herm\bY}.
\end{eqnarray}
%where we can see that the computation is greatly simplified.
%$\bd_x=\text{diag}(\bD)$. Know that $\vec\bW_X=\lambda^{-1} \bU_X^{-1}$, then we have
%\begin{eqnarray}
%	\vec\bU_X^{-\frac{1}{2}}=\lambda^{\frac{1}{2}}\vec\bW_X=\lambda^{\frac{1}{2}}\bC_X \bD_X^{\frac{1}{2}} \bC_X
%\end{eqnarray}
For convenience, we rewrite the unitary transformed pseudo observation model as
\begin{eqnarray}
%	\lambda^{\frac{1}{2}}\bC_X \bD_X^{\frac{1}{2}} \bC_X=\lambda^{\frac{1}{2}}\bC_X \bD_X^{\frac{1}{2}} \bC_X + \bW_X \\
	{\underbrace{\bD_X^{-\frac{1}{2}}\bC_X\herm\hat\bA\herm\bY}_{\bR_X}}
	=\underbrace{\bD_X^{-\frac{1}{2}}\bC_X\herm}_{\bPhi_X}\bX+
	\boldsymbol{\Omega'_X}. \label{eq:pseudoX4}
\end{eqnarray}
The above leads to Lines 1-3 of UAMP-MF in Algorithm 2.
%based on which the UAMP algorithm can be employed.
%Then we have $\bR_X=\bPhi_X\bX+\lambda^{-\frac{1}{2}}\bW_X$, where $\bR_X\triangleq \bD^{\frac{1}{2}}\bC\herm\hat\bA\herm\bY$ and $\bPhi_X\triangleq \bD^{\frac{1}{2}}\bC\herm$.
%Note that $\bD$ is a diagonal matrix, so the calculation of the matrix $\bD^{-\frac{1}{2}}$ is trivial. It is worth mentioning that, no matrix inversion is required and only an EVD is involved.  %The UAMP-MF algorithm is summarized in Algorithm \ref{alg1}.

%the MMSE estimation can be efficiently performed based on the pseudo model in \eqref{eq:pseudoX4} and the prior $p(\bX)$,
\rew{Due to the prior $p(\bX)$, the use of exact $q(\bX)$ often makes the message update intractable. Following (U)AMP, we perform the minimum mean squared error (MMSE) estimation based on the pseudo observation model \eqref{eq:pseudoX4} with prior $p(\bX)$, i.e., project it to be Gaussian.}
%we project it to be Gaussian, which can also be interpreted as performing the minimum mean squared error (MMSE) estimation based on the pseudo observation model \eqref{eq:pseudoX4} with prior $p(\bX)$.
These correspond to Lines 4-11 of UAMP-MF. \rew{Noting that} the prior $p(\bX)$ is separable, i.e., $p(\bX)=\prod_{n,l}p(x_{nl})$, the operations in Lines 10 and 11 are element-wise, \rew{i.e., the function $\bG_X(\bQ_X, V_{\bQ_X})$ is an element-wise function with each entry same as \eqref{g_x} (similarly, $\bG'_X(\bQ_X, V_{\bQ_X})$ denotes its derivative). This is} explained as follows. Due to the decoupling of (U)AMP, we assume the following scalar pseudo models
\begin{equation}
	q_{nl}=x_{nl}+w_{nl}, n=1,...,N, l=1,...,L, \label{eq:scalarmodel}
\end{equation}
where $q_{nl}$ is the $(n,l)$th element of $\bQ_X$ in Line 9 of the UAMP-MF algorithm, $w_{nl}$ represents a Gaussian noise with mean zero and variance $v_{nl}$, and $v_{nl}$is the $(n,l)$th element of $\bV_{Q_x}$ in Line 8 of the UAMP-MF algorithm. This is significant as the complex estimation is reduced to much simpler MMSE estimation based on a number of scalar models \eqref{eq:scalarmodel} with prior $p(x_{nl})$. With the notations in Lines 10 and 11, for each entry $x_{nl}$ in $\bX$, the MMSE estimation leads to a Gaussian distribution
\begin{eqnarray}
\tilde q(x_{nl})=\N(x_{nl};\hat x_{nl},v_{x_{nl}}),
\end{eqnarray}
where $\hat x_{nl}$ and $v_{x_{nl}}$ is the $(n,l)$th element of $\hat\bX$ in Line 11 and the $(n,l)$th element of $\bXi_X$ in Line 10, respectively. We can see that each element $x_{nl}$ has its own variance. To facilitate subsequent processing, \rew{we make an approximation by performing an average operation to each row of $\bXi_X$, i.e., replacing the entries of each row of $\bXi_X$ by their average.} %assuming that the \rew{entries} of each row in $\bX$ have the same variance, which is the average of their variances.
Then \rew{$\{\tilde q(x_{nl})\}$} are collectively characterized by a matrix normal distribution, i.e., $q(\bX)=\MN(\bX; \hat\bX, \bU_X, \bV_X)$ with $\bU_X=\text{diag}(\text{mean}(\bXi_X,2))$ and $\bV_X=\bI_L$, where  $\text{mean}(\bXi_X,2)$ represents the average operation on the rows of $\bXi_X$. This leads to Line 12 of UAMP-MF in algorithm 2.

%The mean matrix and variance matrix can be stacked as $\hat\bX=[x_{n,l}], \forall{n,l}$ and $\bSigma_X=[v_{x_{n,l}}], \forall{n,l}$. The distribution is approximated as $q(\bX)=\MN(\bX; \hat\bX, \bU_X, \bV_X)$ with $\bU_X=\text{diag}(\text{mean}(\bSigma_X,2))$ and $\bV_X=\bI_L$, where  $\text{mean}(\bSigma_X,2)$ represents the average operation on the rows of $\bSigma_X$.

%Then we can obtain the distributions for $\{x_{nl}\}$ (which is approximated to be independent)

\subsubsection{Update of $q(\bA)$}
%Assume the mean and covariance matrices of $\bX$, i.e., $\hat\bX$, $\bU_X$ and $\bV_X$ are available in last iteration,
With the updated $q(\bX)=\MN(\bX; \hat\bX, \bU_X, \bV_X)$ and $q(\lambda)$, we compute the message from $f_Y$ to $\bA$ according to VI. Regarding the message, we have the following result.

\vspace{0.1 in}
\textbf{Proposition 2}: The message from $f_Y$ to $\bA$ can be expressed as a matrix \rew{Gaussian} distribution, i.e., %about $\bA$, which can be expressed as
\begin{equation}
	m_{f_Y\to H}(\bA)\propto \MN(\bA;\overline\bA,\bI_M, \hat\lambda^{-1} \overline\bV_A), \label{eq:msg_fyH_result}
\end{equation}
with
\begin{equation}
	\overline\bA =\bY\hat\bX\herm\overline\bV_A. \label{eq:msg_fyH_mean}
\end{equation}
and
\begin{equation}
	\overline\bV_A = \left(\hat\bX\hat\bX\herm+\Tr(\bV_X)\bU_X\right)^{-1}, \label{eq:msg_fyH_V}
\end{equation}
where
$\hat \lambda=\int_{\lambda} \lambda q(\lambda)$. % and $\hat\bX$ are respectively the means of $\lambda$ and $\bX$, which are computed based on $q(\lambda)$ and $q(\bX)$ in the previous iteration.

\begin{proof}
	See Appendix B.
\end{proof}

Then we combine the message $m_{f_Y\to H}(\bA)$ with the prior of $\bA$ to update $q(\bA)$, \rew{which can also} be realized with UAMP through a whitening operation. The procedure is similar to that for $q(\bX)$, and the difference is that the pseudo model is established row by row (rather than column by column) by considering the form of message $m_{f_Y\to H}(\bA)$.

With the message $m_{f_Y\to H}(\bA)$ and \rew{noting that}
\begin{equation}
\overline\bA= [\overline\ba_1, ..., \overline\ba_M]\tra
%\left(
%\begin{matrix}
%\overline\ba_1\herm \\
%\vdots \\
%\overline\ba_M\herm
%\end{matrix}
%\right),
\end{equation}
where $\overline\ba_m \herm\in\mR^{1\times N}$ is  the $m$th row vector of $\overline\bA$, we have the pseudo observation model
\begin{eqnarray}
\overline\ba_m=\ba_m+\be_m \label{eq:pseudoH1},
\end{eqnarray}
where $\be_m\sim\N(\be_m;0, \hat \lambda^{-1} \overline\bV_A)$.
\rew{Performing} whitening operation to \eqref{eq:pseudoH1} leads to
\begin{eqnarray}
\overline\bV_A^{-\frac{1}{2}}\overline\ba_m =\overline\bV_A^{-\frac{1}{2}}\ba_m+ \bw_m ,
\end{eqnarray}
where $\bw_m = \overline\bV_A^{-\frac{1}{2}} \be_m$, which is white and Gaussian with covariance $\hat\lambda^{-1}\bI$.
Collecting all rows and representing them in matrix form, we have
\begin{eqnarray}
%\cev\bA \cev\bV_A^{-\frac{1}{2}}&=&\bA \cev\bV_A^{-\frac{1}{2}}+ \bW_A \nonumber\\
\overline\bV_A^{-\frac{1}{2}}\overline\bA\herm = \overline\bV_A^{-\frac{1}{2}}\bA\herm+ \boldsymbol{\Omega}_A. \label{eq:pseudoH2}
\end{eqnarray}
UAMP is then performed based on model \eqref{eq:pseudoH2}. Using the idea for updating $q(\bX)$, we obtain the unitary transformed model efficiently.
We first perform an EVD to matrix $\overline \bV_A^{-1}=\overline\bW_A=\hat\bX\hat\bX\herm+\Tr(\bV_X)\bU_X$, i.e.,
\begin{eqnarray}
[\bC_A,\bD_A]=\text{eig}(\overline\bW_A).
\end{eqnarray}
Then the unitary transformed model is given as
\begin{eqnarray}
\bR_A\herm=\bPhi_A\herm \bA\herm+\boldsymbol{\Omega}'_A
\end{eqnarray}
where
\begin{eqnarray}
\bPhi_A &=& \bD_A^{-\frac{1}{2}}\bC_A\herm \\
\bR_A &=& \bD_A^{-\frac{1}{2}}\bC_A\herm\hat\bX\bY\herm,
\end{eqnarray}
and $\boldsymbol{\Omega_A}'$ is Gaussian and white. The above leads to Lines 13-15 of UAMP-MF in Algorithm 2.

Following UAMP in Algorithm 1, we obtain $q(\bA)$ and project it to be Gaussian, which correspond to Lines 16-23 of UAMP-MF Algorithm 2. \rew {The function $\bG_A(\bQ_A, V_{\bQ_A})$ is an element-wise function with each entry similar to \eqref{g_x}. In addition,} similar to the case for updating $q(\bX)$, %each element $h_{mn}$ has a variance.
to accommodate $\{q(h_{mn})\}$ with a matrix \rew{Gaussian} distribution, we make an \rew{approximation by performing an average operation to each column of $\bXi_A$, replacing the entries of each column in $\bXi_A$ by their average.} % assuming the entries in each column of $\bA$ share a common variance, which is the average of their variances.
Then $q(\bA)=\MN(\bA;\hat\bA,\bU_A,\bV_A)$ with $\bV_A=\text{diag}(\text{mean}(\bXi_A,1))$ and $\bU_A=\bI_M$ (i.e., Line 24 of Algorithm 2), where $\text{mean}(\bXi_A,1)$ represents the average operation on the columns of $\bXi_A$.

%The distribution $q(h_{mn})$ is approximated to be Gaussian, i.e.,
%\begin{eqnarray}
%q(h_{mn})=\N(h_{mn};\hat h_{mn},v_{h_{mn}}). \label{eq:pseudoH3}
%\end{eqnarray}
%The  mean matrix and variance matrix can be stacked as  $\hat\bA=[h_{mn}]$ and $\bSigma_A=[v_{h_{mn}}]$, $\forall{m,n}$. The distribution can be approximated as $q(\bA)=\MN(\bA;\hat\bA,\bU_A,\bV_A)$ with $\bV_A=\text{diag}(\text{mean}(\bSigma_A,1))$ and $\bU_A=\bI_M$, where $\text{mean}(\bSigma_A,1)$ represents the average operation on the rows of $\bSigma_A$.

\subsubsection{Update of $q(\lambda)$}
With $q(\bA)=\MN\left(\bA; \hat\bA, \bV_A, \bI_M \right)$ and $q(\bX)=\MN\left(\bX; \hat\bX, \bI_L, \bU_X\right)$, we compute the message from $f_Y$ to $\lambda$, as shown in Proposition 3.  %denoted by $m_{f_Y\to \lambda}(\lambda)$.

%update $q(\lambda)$, i.e., combining the message $m_{f_Y\to \lambda}(\lambda)$ from $f_Y$ to $\lambda$ and the prior $p(\lambda)$. %Regarding this, we have the following result.

\vspace{0.1 in}
\textbf{Proposition 3}: The message from $f_Y$ to $\lambda$ can be expressed as
\begin{eqnarray}
	m_{f_Y\to \lambda}(\lambda)\propto\lambda^{ML}\exp\Big(-\lambda C \Big), \label{eq:msg_fy_lambda}
\end{eqnarray}
where
%\begin{eqnarray}
%	&&\!\!\!\!\!\!C=\Tr\Big(\big(\bY-\hat\bA\hat\bX\big)\herm\big(\bY-\hat\bA\hat\bX\big)\Big)
%	+\Tr\Big(\hat\bX\hat\bX\herm\Tr(\bU_A)\bV_A\nonumber\\
%	&&\ \ +\Tr(\bV_X)\bU_X\hat\bA\herm\hat\bA+\Tr(\bV_X)\bU_X\Tr(\bU_A)\bV_A\Big).
%\end{eqnarray}
\begin{eqnarray}
&& \!\!C=\| \bY-\hat\bA\hat\bX\|^2+M\Tr\Big(\hat\bX\hat\bX\herm\bV_A\Big)\nonumber\\
&& \ \ \ \ \ \ \ \ +L \Tr(\bU_X\hat\bA\herm\hat\bA)+ML\Tr(\bU_X \bV_A). \label{eq:ComputeC}
\end{eqnarray}

\begin{proof}
See Appendix C. 	
\end{proof}

We then combine the message $m_{f_Y\to \lambda}(\lambda)$ and the prior $p(\lambda)\propto 1/\lambda$ to update $q(\lambda)$, i.e.,
\begin{equation}
	q(\lambda)\propto m_{f_Y\to \lambda}(\lambda)p(\lambda) = \lambda^{ML-1}\exp\Big(-\lambda C \Big),
\end{equation}
which is a Gamma distribution.
Then the mean of $\lambda$ is obtained as
\begin{eqnarray}
\hat\lambda = {\int_{\lambda} \lambda q(\lambda )}
={ML}/{C}, \label{eq:lambda}
\end{eqnarray}
which is Line 25 of UAMP-MF.

\subsubsection{Discussion and Computational Complexity}

Regarding UAMP-MF in Algorithm 2, we have the following remarks:
\begin{itemize}
	\item We do not specify the priors of $\bA$ and $\bX$ in Algorithm 2. The detailed implementations of Lines 10, 11, 22 and 23 of the algorithm depends on the priors in a concrete scenario. %Various examples are provided in Section IV.
	\item As the MF problem often has local minima, we can use the strategy of restart to mitigate the issue of being stuck at local minima. In addition, the iterative process can be terminated based on some criterion, e.g., the normalized difference between the estimates of two consecutive iterations is smaller than a threshold.
	\item There are often hyper-parameters in the priors of $\bA$ and $\bX$. If we do not have knowledge on these hyper-parameters, their values need to be leaned or tuned automatically. Thanks to the factor graph and message passing framework, these extra tasks can be implemented by extending the factor graph in Fig. 1. \rew{Some details are provided in next subsection.}
	\item We see that UAMP-MF in Algorithm \ref{UTAMPv1} involves matrix multiplications and two EVDs per iteration, which dominate its complexity. The complexity of the algorithm is therefore in a cubic order, which is low in a MF problem (the complexity of two matrices product is cubic)  %\rew{Extensive examples in Section IV show that UAMP-MF is much faster than many state-of-the-art algorithms in many scenarios.}
	\item \rew{We can also see that the space cost of the algorithm is in the same order of the size of the MF problem, which is reasonable.}
\end{itemize}

\subsection{\rew{Hyper-Parameter Learning}}

\rew{We take MF with a sparse factor matrix as example to show how to learn the hyper-parameters in the priors by taking advantage of the factor graph and message passing techniques.}
\rew{We assume that $\bX$ is sparse and its sparsity rate is unknown. In this case, we can employ the sparsity inducing hierarchical Gaussian-Gamma prior, which is used in SBL \cite{Tipping}} \footnote{\rew{Other sparsity inducing priors may also be used, such as the spike and slab prior \cite{spslab} and horseshoe prior to achieve better robustness \cite{horseshoe}.}}.
%The prior is then $p(x_{nl},\gamma_{nl})= p(x_{nl}|\gamma_{nl})p(\gamma_{nl})$, where
Hence, $p(x_{nl}|\gamma_{nl})=\N(x_{nl};0,\gamma_{nl}^{-1})$ and $p(\gamma_{nl})=Ga(\gamma_{nl}; \epsilon, \eta)$ with $\epsilon$ and $\eta$ being the shape parameter and scale parameter. % of the Gamma distribution, respectively.
While the precision $\gamma_{nl}$ is to be learned, the values for $\epsilon$ and $\eta$ are often set empirically, e.g., $\epsilon= \eta=0$ \cite{Tipping}. It is worth mentioning that $\epsilon$ can also be tuned automatically to improve the performance \cite{UAMPSBL}.

The factor graph representation for this part is shown in Fig. 2. Next, we show the message updates in this sub-graph.  Due to the decoupling of UAMP as shown by the pseudo model \eqref{eq:scalarmodel}, the incoming message to the factor graph in Fig. 2 is Gaussian, i.e.,
\begin{equation}
m_{x_{nl}\to f_{x_{nl}}}(x_{nl})=\mathcal{N} (x_{nl}; q_{nl}, v_{nl}).
\end{equation}
We perform inference on $x_{nl}$ and $\gamma_{nl}$, which can also be achieved by using VI with a variational distribution $q(x_{nl}, \gamma_{nl})= q(x_{nl})q(\gamma_{nl})$.

\tikzstyle{factornode} = [draw, fill=white, circle, inner sep=1pt,minimum size=0.4cm]
\tikzstyle{funnode} = [draw, rectangle,fill=black!100, minimum size = 0.4cm]
\begin{figure}[!t]%[htbp]
	\centering
	\begin{tikzpicture}[auto, node distance=0.8cm,>=latex']
		\node (X)[factornode,] at (0,0) {};
		\node (PXGamma)[funnode, right=0.8cm of X] {};
		\node (Gamma)[factornode, right=0.8cm of PXGamma] {};
		\node (PGamma)[funnode, right=0.8cm of Gamma] {};
		
		\node [above = 0.1cm of X] {$x_{nl}$};
		\node [above = 0.1cm of PXGamma] {$f_{x_{nl}}$};
		\node [above = 0.1cm of Gamma] {$\gamma_{nl}$};
		\node [above = 0.1cm of PGamma] {$f_{\gamma_{nl}}$};
		
		\draw ($(X.west)+(-1.5, 0)$) --node[below = 0.1cm of X]{$\mathcal{N}(x_{nl}; {q}_{x_{nl}}, v_{x_{nl}})$} (X.west);
		\draw (X) -- (PXGamma) -- (Gamma);
		\draw (Gamma) -- (PGamma);
		
		\draw [ ->, >=stealth] ($(X.east)+(0.2, -0.5)$) -- ($(X.west)+(1.2, -0.5)$);
	\end{tikzpicture}
	\centering
	\caption{Factor graph for hyper-parameters learning, \rew{where $f_{x_{nl}}\triangleq p(x_{nl}|\gamma_{nl})$ and $f_{\gamma_{nl}}\triangleq p(\gamma_{nl})$ }.}
	\label{fig:FactorGraph2}
\end{figure}
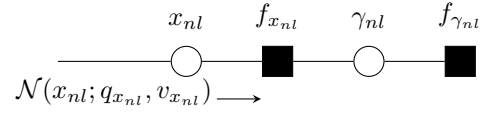

According to VI,
\begin{eqnarray}
	q(x_{nl})&\propto &m_{x_{nl}\to f_{x_{nl}}}(x_{nl}) m_{f_{x_{nl}}\to x_{nl}}(x_{nl})\nonumber\\
	&\propto&\N(x_{nl};\hat x_{nl}, v_{x_{nl}})
\end{eqnarray}
with
$m_{f_{x_{nl}}\to x_{nl}}(x_{nl})$ shown in \eqref{eq:addvi}, and
\begin{eqnarray}
	v_{x_{nl}}=\frac{v_{q_{nl}}}{1+\hat\gamma_{nl}v_{q_{nl}}}, ~
	\hat x_{nl}=\frac{\hat q_{nl}}{1+\hat\gamma_{nl}v_{q_{nl}}}.
\end{eqnarray}
The message $m_{f_{x_{nl}}\to \gamma_{nl}}(\gamma_{nl})$ can be expressed as
\begin{eqnarray}
	m_{f_{x_{nl}}\to \gamma_{nl}}(\gamma_{nl})&\propto& \exp\big( \int_{x_{nl}} q(x_{nl}) \log \mathcal{N}(x_{nl};0,\gamma_{nl}^{-1})\big), \nonumber \\
 &\propto& \gamma_{nl}\exp\big(-{\gamma_{nl}}(|\hat x_{nl}|^2+v_{x_{nl}})\big).
\end{eqnarray}
The message
%$m_{f_{\gamma_{nl}}\to \gamma_{nl}} (\gamma_{nl})$ is the predefined Gamma distribution, i.e.,
$m_{f_{\gamma_{nl}}\to \gamma_{nl}} (\gamma_{nl}) \propto \gamma_{nl}^{\epsilon -1}\exp(-\eta\gamma_{nl})$. According to VI,
\begin{eqnarray}
	q(\gamma_{nl}) &\propto& m_{f_{\gamma_{nl}}\to \gamma_{nl}} (\gamma_{nl}) m_{f_{x_{nl}}\to \gamma_{nl}} (\gamma_{nl}) \nonumber \\
	&\propto& \gamma_{nl}^{\epsilon} \exp\big(-{\gamma_{nl}}(|\hat x_{nl}|^2+v_{x_{nl}}+\eta)\big).
\end{eqnarray}
Thus
\begin{eqnarray}
	m_{f_{x_{nl}}\to x_{nl}} (x_{nl}) &\propto& \exp\big( \int_{\gamma_{nl}} q(\gamma_{nl}) \log \mathcal{N}(x_{nl};0,\gamma_{nl}^{-1})\big), \nonumber \\
	&\propto& \mathcal{N} (x_{nl}; 0, \hat \gamma_{nl}^{-1}), \label{eq:addvi}
\end{eqnarray}
where
\begin{eqnarray}
	\hat\gamma_{nl}=\frac{1+ \epsilon}{ \eta+v_{x_{nl}}+|\hat x_{nl}|^2}.
\end{eqnarray}
The above computations for all the entries of $\bX$ can be collectively expressed as
\begin{eqnarray}
&&\bXi_X=\bV_{Q_X}./\left(\mathbf{1}+\hat\bGamma\cdot\bV_{Q_X}\right), \label{eq:SBL1}\\
&&\hat\bX={\hat\bQ_X}./\left(\mathbf{1}+\hat\bGamma\cdot\bV_{Q_X}\right), \\
&&\hat\bGamma=(\mathbf{1}+ \epsilon \mathbf{1})./\left( \eta \mathbf{1}+\bXi_X+|\hat\bX|^{.2}\right),\label{eq:SBL2}
\end{eqnarray}
where $\hat\bGamma$ is a matrix with $\{\gamma_{nl}\}$ as its entries.

% As algorithms require different In next section, we will use extensive examples to demonstrate the

%Using the above steps, we can observe that, there is only a eigenvalue decomposition for each iteration.

%\begin{table}[!hbp]
%	\centering
%	\caption{Distributions and factors in \eqref{eq:factor1}}
%	\begin{tabular}{lll}
%		\hline
%		Factor & Distribution & Function\\
%		\hline
%		$f_{\br}$ & $p\left(\br|\bz,\beta\right)$ & $\N\left(\bz;\br,\beta^{-1}I\right)$\\
%		$f_{\bz}$ & $p\left(\bz|\bx\right)$ & $\delta\left(\bz-\bPhi \bx\right)$ \\
%		$f_{\bx}$ & $p(\bx|\bc,\bb)$ & $\delta\left(\bx-\bb\otimes\bc\right)$\\
%		$f_{x_{n,k}}$ & $p\left(x_{n,k}|b_k,c_{n}\right)$ & $\delta\left(x_{n,k}-b_k c_{n}\right)$ \\
%		$f_{\bc}$ & $p(\bc)$ & prior of $\bc$, e.g., prior promoting sparsity \\ %SBL sparse prior
%		$f_{\bb}$ & $p(\bb)$ & prior of $\bb$\\ %$\N(\bb;0,I)$
%		$f_{\beta}$ & $p(\beta)$ & $k\beta^{-1}$ \\
%		\hline
%	\end{tabular}
%\end{table}

\section{UAMP-MF for Integrated Near Field Localization and Communications}

In this section, we apply the developed UAMP-MF algorithm to tackle the JNFLSD problem. It is noted that the application is not straightforward as the matrix $\bA$ has a structure with its columns given as steering vectors parameterized by $\{d_k,\theta_k\}$. 

An advantage of integrating UAMP with VI in developing the UAMP-MF algorithm is that it facilitates the reinforcement of the structure of $\bA$ thanks to the decoupling of UAMP, i.e., by executing Lines \ref{alg:UAMPMF_VQA}-\ref{alg:UAMPMF_QA}, which produce $\bV_{Q_A}$ and $\bQ_A$, we have the following pseudo model
\begin{equation}
	\bq_{A,z}= \ba(d_z,\theta_z) + \bw_z,   z=1, ... Z, \label{eq:pse_qa}
\end{equation}
where $\bq_{A,z}$ is the $z$th column of $\bQ_A$, and $\bw_z$ is approximated to a white noise with mean zero and variance $\sigma^2=\left<\bV_{Q_A}\right>$. Next, we estimate $d_z$ and $\theta_z$ based on the above model. To simplify the notations, we drop the subscript $z$ in this section.

The challenge lies in the non-linearity of $\ba(d,\theta)$ on the parameters $d$ and $\theta$. To overcome the challenge, we adopt a dynamic linearization method detailed in the following. As the algorithm is an iterative one, where the estimates of $d$ and $\theta$ in the last round of iteration, denoted by $d'$ and $\theta'$, are available. Then the partial derivative vector $\be_\theta(d',\theta')$ and $\be_d (d', \theta')$ can be expressed as
\begin{eqnarray}
	\be_\theta(d',\theta')= \frac{\partial \ba(d,\theta)}{\partial \theta}\large|_{d=d' ,\theta=\theta'}=\left[e_{\theta_1},...,e_{\theta_R}\right]\tra , \\
	\be_d(d',\theta')= \frac{\partial \ba(d,\theta)}{\partial d}\large|_{d=d', \theta=\theta'}
	=\left[e_{d_1},...,e_{d_R}\right]\tra,
\end{eqnarray}
where
\begin{eqnarray}
	&&\!\!\!\!\!\!\!e_{\theta_r}\triangleq\frac{\partial a_r(d,\theta)}{\partial \theta}\large|_{d=d' ,\theta=\theta'}=a_r(d',\theta')\nonumber\\
	&&\times\left(\frac{j2\pi}{\lambda}d b_r \sin\theta'\sqrt{d'^2+b_r^2+2d'b_r\cos\theta'}\right) \\
	&&\!\!\!\!\!\!\!e_{d_r}\triangleq\frac{\partial a_r(d,\theta)}{\partial d}\large|_{d=d' ,\theta=\theta'}=a_r(d',\theta')\nonumber\\
	&& \times \frac{-j2\pi}{\lambda}\Big(\sqrt{d'^2+b_r^2+2d'b_r\cos\theta'}(d'+b_r \cos\theta')-1\Big). \nonumber \\
\end{eqnarray}
Hence, with the first-order Taylor expansion, the vector $\balpha(d,\theta)$ can be approximately linearized at $(d',\theta')$ as
\begin{eqnarray}
	\ba(d,\theta)\!\!\!\!\!&\approx& \!\!\!\!\! \ba(d',\theta')+ \be_\theta(d',\theta')(\theta-\theta') +\be_d(d',\theta') (d-d')\nonumber\\
	&=&\ba'+ \be_\theta'(\theta-\theta') +\be'_d(d-d'), \label{eq:pse_taylor_a}
\end{eqnarray}
where $\ba'\triangleq\ba(d',\theta')$, $\be_\theta'\triangleq\be_\theta(d',\theta')$ and $\be_d'\triangleq\be_d(d',\theta')$.

According to \eqref{eq:pse_qa} and \eqref{eq:pse_taylor_a}, we have
\begin{equation}
	\bxi_z=\bq_{A,z}-\ba'=
	%\left(\be^t_{d_k}, \be^t_{\theta_k} \right)
%	\left(
%	\begin{array}
%		{c}
%		d_k^{t+1}-d_k^{t}\\
%		\theta_k^{t+1}-\theta_k^{t}
%	\end{array}
%	\right) +  \bw_z
	\bE_z \bc_z + \bw_z, \label{eq:LS_xi}
\end{equation}
where
\begin{equation}
	\bE =\left[\be_{d}, \be_{\theta} \right],
\end{equation}
\begin{equation}
	\bc=\left[
		\begin{array}
		{c}
		d-d'\\
		\theta-\theta'
	\end{array} \right].
\end{equation}
%where $\sigma=\left<\bV_{Q_A}\right>$,
Then, with model \eqref{eq:LS_xi}, we perform a LS estimation of $\bc$, which is equivalent to the MMSE estimation with a priori mean $\bc^{a}=0$ and an inverse of covariance matrix $\bV_c^a=0$, i.e.,
\begin{eqnarray}
\bV_{c}&=&(\frac{1}{\sigma^2}\bE^H\bE)^{-1},  \label{eq:LS_Vc}\\
	\hat\bc&=&\sigma^{-2}\bV_{c}\bE^H\bxi. \label{eq:LS_c}
\end{eqnarray}
Then we can get the updated estimates of $d$ and $\theta$ as
\begin{equation}
	\left[
	\begin{array}
		{c}
		\hat d\\
		\hat \theta
	\end{array} \right] = \hat\bc + \left[
\begin{array}
{c}
d'\\
\theta'
\end{array} \right].
\end{equation}
For the LS estimation, we have the following remarks:
\begin{itemize}
	\item As the LS estimation is a special case of the MMSE estimation, the matrix computed in (66) is actually the a posteriori covariance matrix of $\bA$, which facilitates the incorporation of the estimation to the UAMP-MF algorithm.
	\item The computational complexity involved in \eqref{eq:LS_Vc} and \eqref{eq:LS_c} is low due to the small size of the matrices, i.e., $\bE^H\bE$ and $\bV_c$ are $2 \times 2$ matrices.
\end{itemize}
%\begin{eqnarray}
%	\hat d=\hat\bc(1)+d_k, \\
%	\theta_k^{t+1}=\hat\bc_k(2)+\theta_k^{t}
%\end{eqnarray}
In the above, we only detail the computations for a single column of $\bQ_A$. After applying it to all the columns, we can get $\{\hat d_z, \hat{\theta}_z\}$.
Then according to the UAMP-MF algorithm, we need to produce the estimate $\hat \bA$ and the covariance matrix $\bXi_A$. With the estimates $\{\hat d_z, \hat{\theta}_z\}$, it is straightforward to obtain that
\begin{equation}
	\hat\bA=\left[\ba(d_1,\theta_1),...,\ba(d_Z,\theta_Z)\right]. \label{eq:HatA}
\end{equation}
For the computation of the covariance matrix, we use the linearized {model \eqref{eq:pse_taylor_a} for each column of $\bA$} and then average them, leading to}
\begin{equation}
\bXi_A=\frac{1}{Z}\sum_z \bE_z \bV_{c_z} \bE_z^H.
\end{equation}

%\begin{equation}
%\bV_A=\text{diag}(\bXi_A).
%\end{equation}

\begin{algorithm}
	\caption{UAMP-MF Based JNFLSD Algorithm}
	{\textbf{Initialization}: $\tilde K=U_{max}$, $\bU_A=\bI_R$,  $\hat\bA=\bA^0$, $\bV_X=\bI_L$.\\
		\textbf{Repeat}
		\begin{algorithmic}[1]
			\STATE Execute Lines \eqref{alg:UAMPMF_X1}-\eqref{alg:UAMPMF_VQx} of UAMP-MF in Algorithm \ref{alg:UAMPMF} to obtain $\bQ_X$ and $\bV_{Q_X}$
			\STATE $\hat\bX=\bQ_X./\left(1+\bV_{Q_X} \cdot \bGamma\right)$
			\STATE $\bXi_X=\bV_{Q_X}./\left(1+\bV_{Q_X} \cdot \bGamma\right)$
			\STATE $\bgamma=(\epsilon+1)\boldsymbol{1}./\left(L^{-1}\big(|\hat\bX|^{.2}+\bXi_X\big)\boldsymbol{1}_{L}\right)$
			\STATE $\epsilon=\sqrt{\log\left(<\bgamma> \right)-<\log\left(\bgamma\right)>}$
			\STATE $\bGamma= \bgamma\boldsymbol{1}_{L}\tra$
			\STATE Execute Lines \eqref{alg:UAMPMF:H1}-\eqref{alg:UAMPMF_QA}  of UAMP-MF in Algorithm \ref{alg:UAMPMF} to obtain $\bV_{Q_A}$ and $\bQ_A$
			\STATE $\forall z:$ Construct $\bE_z$ and $\bxi_z$ according to \eqref{eq:LS_xi}.
			\STATE $\forall z: \bV_{c_z}=\sigma\left(\bE_z^H\bE_z\right)^{-1}$
			\STATE $\forall z: \hat\bc_z=\sigma^{-1}\bV_{c_z}\bE_k^H\bxi_z$
			\STATE $\forall z: [d_z^{t+1}, \theta_z^{t+1}]\tra= \hat\bc_z + [d_z^{t}, \theta_z^{t}]\tra$
			\STATE $\bXi_A=\frac{1}{Z}\sum_z \bE_z \bV_{c_z} \bE_z^H$
			\STATE $\hat\bA=\left[\ba(d_1,\theta_1),...,\ba(d_Z,\theta_Z)\right]$ \label{alg:UAMP_AatH}
			\STATE Execute Lines \eqref{alg:UAMPMF_VH}-\eqref{alg:UAMPMF_C} of UAMP-MF in Algorithm \ref{alg:UAMPMF}
		\end{algorithmic}
		\textbf{Until terminated}\\
		Output $[d_k, \theta_k]$ and perform differential demodulation.
		\label{alg:UACESD}}
\end{algorithm}

We note that the number of candidate columns in $\hat\bA$ can be often larger than $K$, i.e., its size is $R \times Z$ , and the corresponding matrix  $\hat\bX$ has a size of $Z \times L$. Regarding this, we note the following.
\begin{itemize}
	\item It is crucial to imposing a sparse prior on $\bX$, so that the columns of $\hat\bA$ with false distance-angle pairs correspond to a (nearly) zero row of $\hat\bX$. In this paper, we adopt the Gauss-Gamma prior.
	\item We further note that the rows of $\hat\bX$ have common support, which should be exploited to achieve better performance. This can be achieved using the Gauss-Gamma priors, where we impose that the elements of a row share a single precision $\gamma_z$. Hence, we have
    \begin{eqnarray}
		p(\bX)&=& \prod_{l} p(\bx_l|\bgamma)p(\bgamma)=\prod_{l,z}p(x_{z,l}|\gamma_z)p(\gamma_z)\nonumber\\
        &=&\prod_{l,z}\CN(x_{z,l};0,\gamma_z^{-1})\Ga(\gamma_z;\epsilon,\eta).
	\end{eqnarray}
	\item It is shown in \cite{UAMPSBL} that, the performance of sparse signal recover can be significantly improved by automatically tuning the hyper parameter $\epsilon$. We adopt the update rule in \cite{UAMPSBL}, i.e., $\epsilon$ is updated with the following equation in each iteration
\begin{equation}
		\epsilon=\sqrt{\log\left(<\bgamma> \right)-<\log\left(\bgamma\right)>},
	\end{equation}
where
\begin{equation}
	\boldsymbol{\gamma}=(\epsilon+1)\boldsymbol{1}./\left(L^{-1}\big(|\hat\bX|^{.2}+\bXi_X\big)\boldsymbol{1}_{L}\right).
\end{equation}
\end{itemize}

The UAMP-MF based JNFLSD algorithm is summarized in Algorithm 3. The algorithm is an iterative one, and the iteration can be terminated when the difference between the estimates of $\bA$ or $\bX$ in two consecutive iterations is less than a threshold. At the end of the iteration, the algorithm outputs distance-angle pairs and the estimate  $\hat\bX$, based on which differential demodulation can be performed to make decision on the transmitted symbols. Regarding the algorithm, we have the following remarks.
\begin{itemize}
\item For the initialization of matrix $\bA$, we propose to use a spatial power spectrum method, where the spatial power spectrum is defined as
\begin{equation}
	\mathcal{S}(d_z, \theta_z)= {\ba (d_z, \theta_z)\herm \bY\bY\herm\ba(d_z, \theta_z)}. %{||\ba(d_z, \theta_z)||^2||\bY||^2} 
	\label{eq:spectrum}
\end{equation}	
Through a coarse scan with the power spectrum, we can determine the areas where users may be located. Then these areas are divided uniformly with the distance-angle pairs $\{d_z, \theta_z\}$, based on which matrix $\bA$ can be initialized.
\item The estimates of the matrices $\bA$ and $\bX$ are updated in each iteration. The rows of $\bX$ with small values and the corresponding columns of $\bA$ are killed, leading to low complexity of the iterative process.
\item Thanks to the simultaneous estimation of $\bA$ and $\bX$, the algorithm does not require the users to sent pilot signals, which makes the transmission efficient and suitable to dealing with dynamic scenarios. It is noted that, in the case that few pilot symbols are available, differential modulation is not needed.
\end{itemize}

%\begin{table}[htb]
%	\centering
%	\begin{minipage}[t]{0.9\linewidth}
%		\caption[table]{Simulation Configurations}
%		\label{tab:para}
%		\begin{tabular*}{\linewidth}{lp{16cm}}
%			\toprule[1.5pt]
%			{Antenna array in BS} & $R=128$~~\\
%			{Number of active users } & $K=1-7$~~\\
%			{Carrier frequency} & $30GHz$~~\\
%			{The distribution of $\theta$} & $U[\frac{\pi}{6},\frac{5\pi}{6}]$~~\\
%			{Maximum Distance} & $L_{max}=10-70$\\
%			{Minimum Distance} & $L_{min}=3$\\
%			\bottomrule[1.5pt]
%		\end{tabular*}
%	\end{minipage}
%\end{table}

\section{Simulation Results}
Various simulation results are provided to verify the performance of the proposed UAMP-MF based JNFLSD algorithm. The system settings are as follows. A uniform linear antenna array with $R=128$ elements and half-wavelength spacing is employed, and the center frequency $f_c=30$GHz, so the Rayleigh distance is about 82 meters. We assume that $K$ active users are uniformly distributed in an area of interest with distance range [5m, $d_{max}$] and angle range [$30^o,150^o$]. In the simulations, we vary the maximum distance $d_{max}$ from 20m to 70m  to investigate the impact of maximum distance on the performance of localization and communications.   %The distances between the BS and users \que{$d_{max}$} range from 10 meters to 70 meters and .  \que{To highlight the impact of the near-field, we ignore the large scale path loss in the channel.} 
We compare the proposed UAMP-MF based JNFLSD algorithm with existing algorithms, including the on-grid SW-OMP algorithm \cite{Rod2018}, where far-field is assumed, the off-grid SIGW-OLS algorithm \cite{Gon2021}, where far-field is also assumed, and the on-grid algorithm NF-SOMP  {\cite{Wei2022}}, \cite{Tropp2005}, where near-field is assumed. The signal to noise power ratio (SNR) is defined as
\begin{equation}
	\mathrm{SNR}=\frac{\Vert{\bA\bX}\Vert^{2} /RKL}{\sigma^2},
\end{equation}  
where the numerator is the power of the received signal per antenna per user, and $\sigma^2$ is the power of noise.
The localization performance is evaluated using the normalized mean squared errors (NMSE) of the distance estimation and the MSE of the angle estimation. 
%which are respectively defined as
%\begin{eqnarray}
%	\text{NMSE}(d)=\frac{1}{TK}\sum_{t,k}10\log \frac{\|\hat d_{t,k}-d_{t,k}\|^2}{\|d_{t,k}\|^2},  \\
%	\text{MSE}(\theta)=\frac{1}{TK}\sum_{t,k}10\log {\|\hat \theta_{t,k}-\theta_{t,k}\|^2},
%\end{eqnarray} 
%where $T$ is number of Monte Carlo simulations, $K$ is the number of active users, and $\hat d_{t,k}$ and $\hat \theta_{t,k}$  are the estimated distance and angle of the $k$th user in the $t$th Monte Carlo simulation, respectively. 
The communication performance is evaluated using the bit error rate (BER) and frame error rate (FER).   

%The simulation configuration is shown in Table I. We compare the proposed UAMP-MF algorithm with the existing methods, including the on-grid far field SW-OMP algorithm \cite{Rod2018}, the off-grid far field SIGW-OLS algorithm \cite{Gon2021} and on-grid near field algorithm NF-SOMP \cite{Cui2022}.

\begin{figure}[!t]
	\centering
	\includegraphics[width=0.5\textwidth]{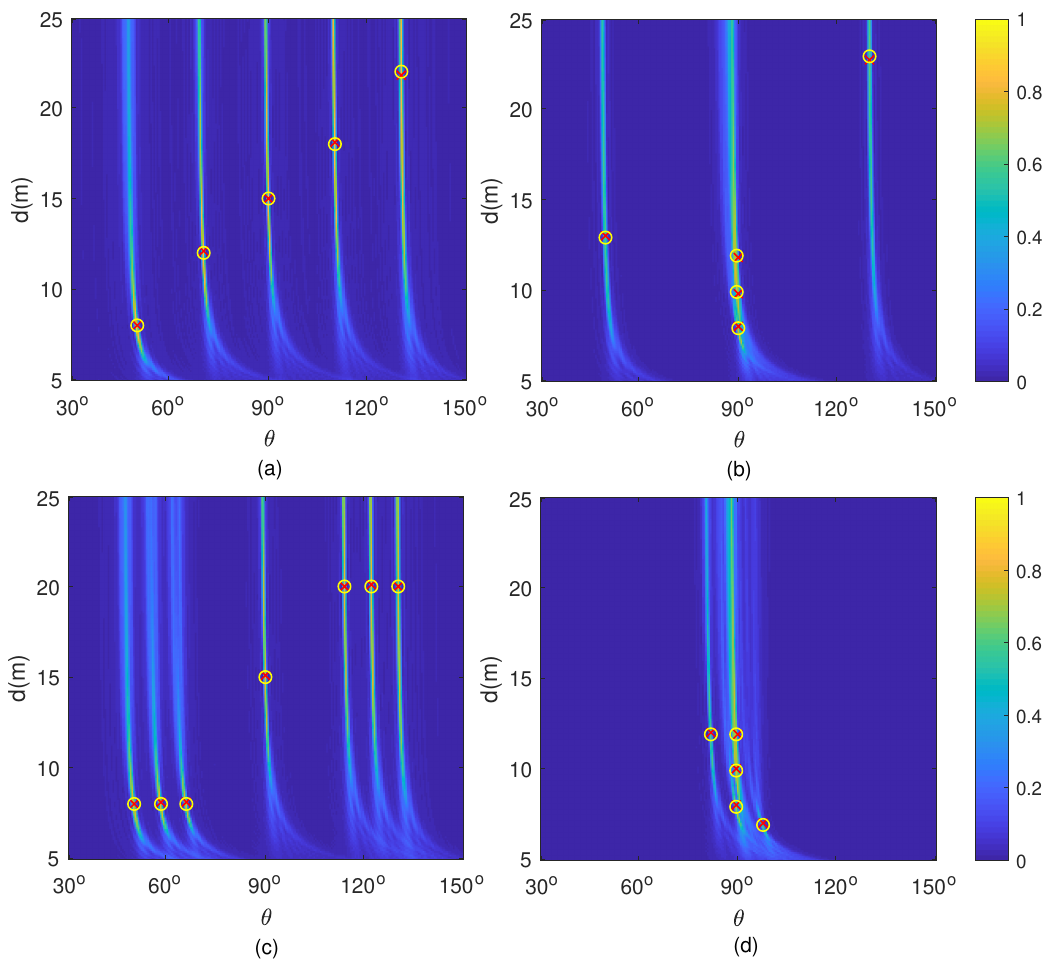}
	\centering
	\caption{Spatial power spectrum (normalized), the true locations (denoted by ``$\mathrm{o}$") and estimated locations (denoted by ``$\times$") of active users using the proposed algorithm.}
	\label{fig:spectrum}
\end{figure}

We first examine the spatial power spectrum defined in \eqref{eq:spectrum} in Fig. \ref{fig:spectrum} for some typical scenarios, where the spectrum is normalized and $SNR=-4dB$. The horizontal axis and vertical axis represent the angle and distance, respectively. We assume that there are 5 or 7 active users located within the area of interest, which are indicated using ``$\mathrm{o}$" in the figure. The estimated locations using the proposed UAMP-MF based algorithm are also shown, which are denoted by ``$\times$" in the figure. In Fig. \ref{fig:spectrum}(a), the users are well separated in both the angle domain and distance domain, which can be distinguished based on the spatial power spectrum. As expected, the UAMP-MF based algorithm can accurately localize all users. In Fig. \ref{fig:spectrum}(b), three users out of five uses locate closely in the distance domain and angle domain, which can be hardly distinguished based on the spatial power spectrum. We can see that the UAMP-MF based algorithm can still estimate their locations accurately. In Fig. \ref{fig:spectrum}(c), there are 7 active users, where two groups of 3 users have the same distances with BS, and locate closely in the angle domain. Again the UAMP-MF based algorithm works well. Fig. \ref{fig:spectrum}(d) shows a challenging case, where all users locate closely in both angle and distance, where we can see that the proposed algorithm still delivers promising performance.            

%We first show in Fig. \ref{fig:spectrum} the normalized spatial power spectrum under the condition of maximum distance $d_max=25$m and $SNR=-4dB$, where the horizontal axis is angle and the vertical axis is distance. The coordinate axes in the figure are divided into $200$ grids respectively, that is, the space region is divided into $Z=4\times 10^{4}$ grid points. \eqref{eq:spectrum} is used to calculate the spatial power spectrum of each grid point $z$. Fig. \ref{fig:spectrum}(a)-(d) shows 4 examples, each with a different number and location of active users. As can be seen from Fig. \ref{fig:spectrum}.(a)(c), when different users have different angles, active users can be distinguished from the spatial spectrum. When users are at the same angle and at different distances, the spatial spectrum will be clustered together and it will be difficult to distinguish. The circles in Fig. \ref{fig:spectrum} represent real active user coordinates, and the red crosses represent the coordinates estimated by the UAMP-MF algorithm. It can be seen that the UAMP-MF algorithm proposed in this paper can estimate user coordinates with high precision.

\begin{figure}[!t]
	\centering
	\includegraphics[width=0.4\textwidth]{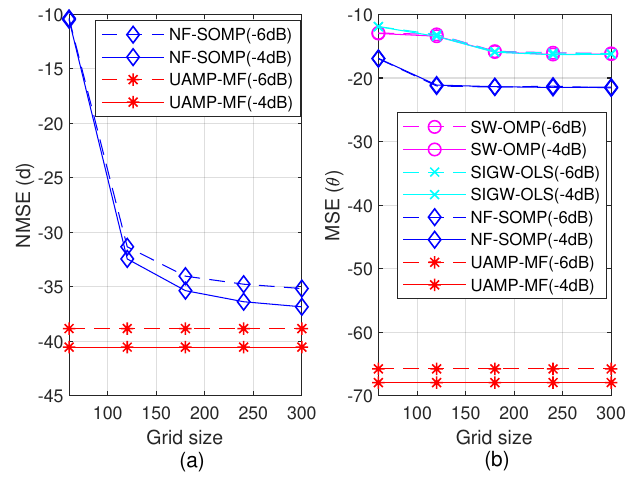}
	\centering
	\caption{(a) NMSE of distance estimation and (b) MSE of angle estimation versus grid size with $d_{max}$= 20m and SNR=-4dB and -6dB.}
	\label{fig:MSE_vsGrid}
\end{figure}

\begin{figure}[!t]
	\centering
	\includegraphics[width=0.4\textwidth]{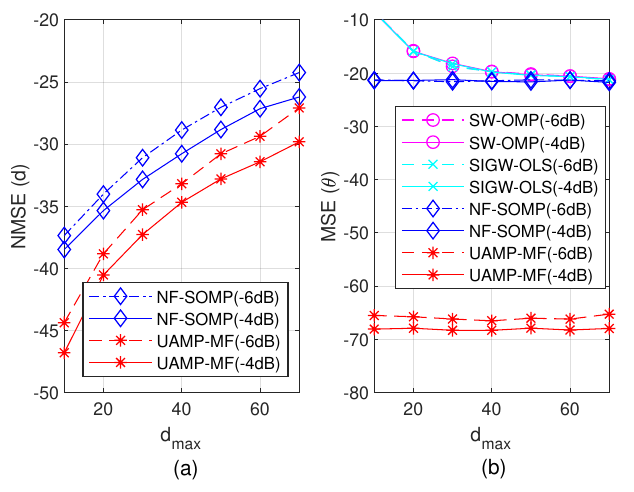}
	\centering
	\caption{(a) NMSE of distance estimation and (b) MSE of angle estimation versus  $d_{max}$ with SNR=-4 and -6dB.}
	\label{fig:MSE_DTvsLmax}
\end{figure}

Next we compare the UAMP-MF based JNFLSD algorithm with  SW-OMP, SIGW-OLS and NF-SOMF. It is noted that SW-OMP, SIGW-OLS and NF-SOMF all use a grid, and the grid size impacts their performance. In contrast, the proposed UAMP-MF based algorithm does not rely on the use of a grid. In addition, SW-OMP and SIGW-OLS assume far-field signal models and do not have the capability of distance estimation. So in evaluating the NMSE performance of distance estimation, SW-OMP and SIGW-OLS are absent. We first investigate the impact of grid size on the performance of the algorithms, and the NMSE of distance estimation and MSE of angle estimation are shown in Figs. \ref{fig:MSE_vsGrid} (a) and \ref{fig:MSE_vsGrid} (b), respectively. In the figures, the horizontal axis represents the number of grid points in angle (for SW-OMP, SIGW-OLS and NF-SOMF) and distance (for NF-SOMF). It can be seen that with the increase of grid points, the performance of SW-OMP, SIGW-OLS and NF-SOMF is improved but with the cost of increased complexity. When the number of grid points is larger than 240, the performance has no significant change. So, in subsequent simulations, for SW-OMP, SIGW-OLS and NF-SOMF, the number of grid points in angle and distance is set to 240, i.e., the grid size is $240 \times 240$. From the results, we can see that the proposed algorithm significantly outperforms other algorithms in both distance estimation and angle estimation. In angle estimation, the performance of SW-OMP and SIGW-OLS is not good due to the model mismatch problem as they use the far-field assumption.

%\begin{figure}[!t]
%	\centering
%	\includegraphics[width=0.45\textwidth]{BERFERvsLmax.pdf}
%	\centering
%	\caption{BER and FER versus maximum distance $d_{max}$.}
%	\label{fig:BERvsLmax}
%\end{figure}

%Fig. \ref{fig:MSE_DTvsLmax} and Fig. \ref{fig:MSE_DTvsSNR} show the curves of normalize mean square error (NMSE) with different maximum distance $d_{max}$ and SNRs respectively. 

%Here we define NMSE$(d)$ and NMSE$(\theta)$ as
%\begin{eqnarray}
%\text{NMSE}(d)=\frac{1}{TK}\sum_{t,k}10\log \frac{\|\hat d_{t,k}-d_{t,k}\|^2}{\|d_{t,k}\|^2} \nonumber\\
%\text{NMSE}(\theta)=\frac{1}{TK}\sum_{t,k}10\log \frac{\|\hat %\theta_{t,k}-\theta_{t,k}\|^2}{\|\theta_{t,k}\|^2} \nonumber
%\end{eqnarray}

Then, we investigate the performance of localization versus the maximum distance $d_{max}$ 
%in terms of the NMSE of distance estimation and MSE of angle estimation, 
and the results are shown in Fig. \ref{fig:MSE_DTvsLmax}.
% (a) and (b), respectively, where SNR=-4 dB and -6dB. Fig. \ref{fig:MSE_DTvsLmax} (a) only includes UAMP-MF and NF-SOMP as SW-OMP and SIGW-OLS assume far-field signal models and do not have the capability of distance estimation.     %\que{The number of active users is $K=3$.} 
We can see that, the proposed algorithm achieves the best performance in all cases. With the increase of the maximum distance $d_{max}$, the performance of distance estimation degrades, which is expected as the signal model tends to be a far-field one, where the contribution of the distance to the distance-angle steering vector diminishes. In contrast, the performance of the angle estimation does not change much with $d_{max}$, which is because in both near-field and far-field, the parameter of angle always plays an important role in the steering vector. It can also be observed from the figure that the performance of angle estimation of SW-OMP and SIG-OLS is improved considerably with the increase of distance. This is because, with the increase of the distance, the mode mismatch due to the far-field assumption is alleviated. When the distance is small, the SW-OMP and SIG-OLS suffer from significant performance loss due to the severe model mismatch.      
%When $d_{max}$ approaches the Rayleigh distance, the performance of SW-OMP and SIG-OLS approaches that of NF-SOMP based on the near-field hypothesis. This phenomenon can be understood as when the $d_{max}$ is small, the error of the far-field model is large, which will lead to large performance loss.}

\begin{figure}[!t]
	\centering
	\includegraphics[width=0.4\textwidth]{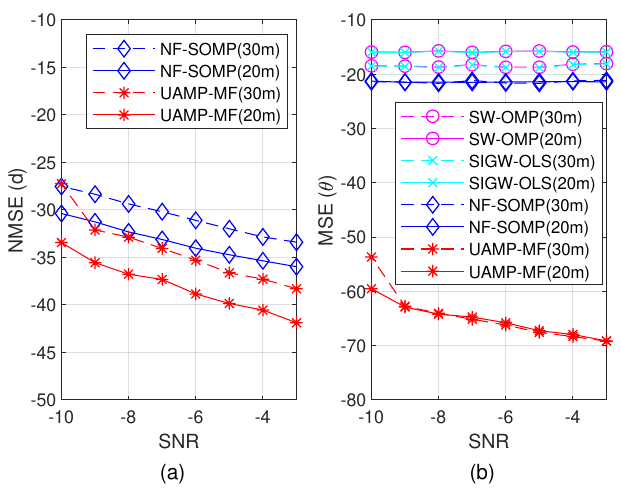}
	\centering
	\caption{(a) NMSE of distance estimation and (b) MSE of angle estimation versus SNR with $d_{max} = 20$m and $30$m.}
	\label{fig:MSE_DTvsSNR}
\end{figure}

\begin{figure}[!t]
	\centering
	\includegraphics[width=0.4\textwidth]{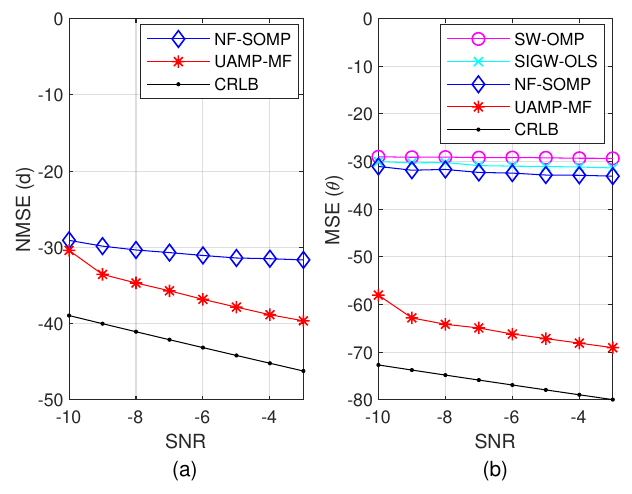}
	\centering
	\caption{ Comparison with CRLB (a) distance estimation and (b) angle estimation.}
	\label{fig:CRLB}
\end{figure}

In Fig. \ref{fig:MSE_DTvsSNR}, we show the NMSE of the distance estimation and MSE of angle estimation versus SNR, where the maximum distance $d_{max}$ =20 and 30 meters. As expected, with the increase of the SNR, the performance of distance estimation becomes better, and the proposed algorithm outperforms NF-SOMP consistently. We can see from  \ref{fig:MSE_DTvsSNR} (b) that the MSE of angle estimation of the proposed algorithm is significantly better than that of other algorithms. In addition, the NMSE performance of angle estimation does not depend too much on the maximum distance $d_{max}$, which is consistent with the results in Fig. \ref{fig:MSE_DTvsLmax}. In addition, the MSE of angle estimation of other algorithms does not improve too much with the SNR, which is because the grid mismatch or model mismatch dominate the errors.      

We compare the performance of the algorithms with the CRLB, which is given in \cite{Grosicki2005}, and the results are shown in Fig. \ref{fig:CRLB}. In the simulations, we assume three users located at  $(5.3 \mathrm{m},60.3^o), (10.3\mathrm{m},90.3^o)$ and $(15.3\mathrm{m},120.3^o)$, respectively. We can see that, compared to other algorithms NF-SOMP, SW-OMP and SIGW-OLS, the proposed algorithm delivers performance much closer to the CRLB.  

%shows the MSE and CRLB \cite{Grosicki2005CRLB} of distance $d$ and angle $\theta$ when the number of users $K=3$ and there polar coordinate positions are fixed as $(5.3m,60.3^o), (10.3m,90.3^o)$ and $(15.3m,120.3^o)$  respectively. It can be seen from the figure that the proposed UAMP-MF method is much closer to CRLB than the existing methods in literature. 
%shows the curves of NMSE$(d)$ and NMSE$(\theta)$ with $d_{max}$ when the number of active users $K=3$ and SNR=-4,-6 and -8dB. It can be seen from Fig. \ref{fig:MSE_DTvsLmax}.(a) that the NMSE$(d)$ deteriorates significantly with $d_{max}$, and the NMSE$(\theta)$ also deteriorates slightly with distance.

%We show in  Fig. \ref{fig:MSE_DTvsSNR}  the curves of NMSE$(d)$ and NMSE$(\theta)$ with SNR under the condition of $K=3$ and maximum distance $d_{max}$=20, 30 and 40 meters. It can be seen from the figure that NMSE$(d)$ and NMSE$(\theta)$ decreases linearly with SNR, but when SNR=-10dB, noise will worsen the estimation performance. By comparison with figure (a) and (b), it can also be found that $d_{max}$ has a greater impact on NMSE$(d)$, and a closer distance $d$ will improve the estimation accuracy of NMSE$(d)$. However, $d_{max}$ has little influence on the accuracy of NMSE$(\theta)$.

\begin{figure}[!t]
	\centering
	\includegraphics[width=0.4\textwidth]{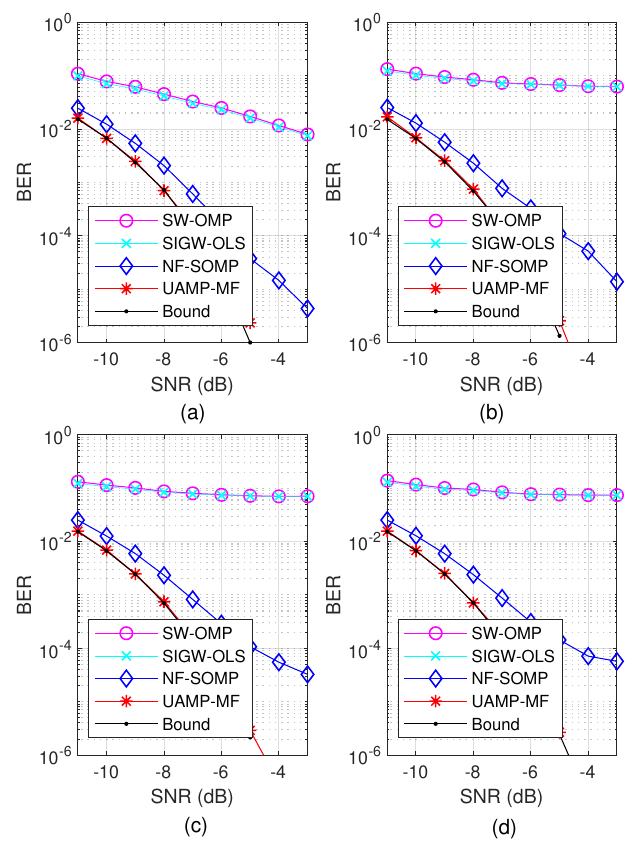}
	\centering
	\caption{BER versus SNR with $K=1,3,5,7$ and $d_{max}$=30m.}
	\label{fig:BERvsSNR}
\end{figure}

%\begin{figure}[!t]
%	\centering
%	\includegraphics[width=0.45\textwidth]{AERvsSNRandBERvsK.pdf}
%	\centering
%	\caption{(a) AER versus SNR with $K=2, 3, 4, 5$, (b) BER versus $K$ with SNR=-6dB.}
%	\label{fig:AERvsSNR}
%\end{figure}

\begin{figure}[!t]
	\centering
	\includegraphics[width=0.4\textwidth]{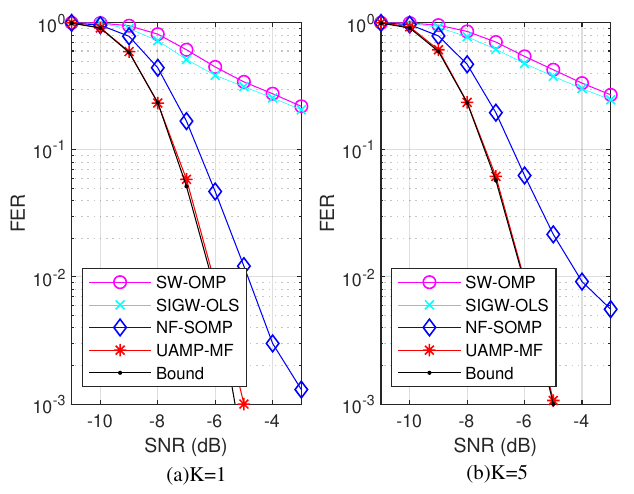}
	\centering
	\caption{FER versus SNR with $K=1$ and $3$ and $d_{max}$=30m. }
	\label{fig:FERvsSNR}
\end{figure}

We compare the BER performance of the proposed algorithm with SW-OMP, SIGW-OLS and NF-SOMP. The results are shown in Fig. \ref{fig:BERvsSNR}, where the number of users $K=1, 3, 5, 7$ in (a), (b), (c) and (d), respectively. The FER performance is shown in Figs. \ref{fig:FERvsSNR} (a) and \ref{fig:FERvsSNR}(b), where the number of users $K=1$ and $3$, respectively. In all cases, the maximum distance $d_{max}$=30m.   
%We note that SW-OMP and SIGW-OLS have a far-field assumption, and NF-SOMP has a near-field assumption.  SW-OMP and NF-SOMP are grid-based methods, \que{where the sensing area is divided into 20 and 30 grids in the range and Angle domains, respectively, and other parameters are completely consistent with the proposed UAMP-MF algorithm.} We also note that, except the proposed method, none of the other methods can estimate of active users, so the number of active users is assumed known. 
In addition, we also show the BER and FER performance bounds, which are obtained by assuming the locations of all users are exactly known, i.e., the matrix $\bA$ is known. From the results we can see that, in all cases, SW-OMP and SIGW-OLS delivers poor BER performance. This is because they make far-field assumption, leading to significant model mismatch. We can also see that NF-SOMP only delivers good performance in the case of $K=1$, and its performance deteriorates with the increase of $K$. This is because NF-SOMP adopts grid-based near-field model, and it works well in the case $K=1$, where there is no inter-user interference. However, in the case of multiple users, NF-SOMP has limited capability to deal with the inter-user interference due to the energy leakage in determine the locations of the users, which results in a worse estimate of $\bA$, thereby a worse estimate of $\bX$.    
%and assumes that the number of active users is known, when $K=1$, NF-SOMP algorithm can accurately find the grid where the user is located. When $K\geq 2$, due to the inherent spectrum leakage problem of griding, the grid where the user is located will have estimation errors, resulting in detection errors.} 
In all cases, the proposed algorithm delivers significantly better BER and FER performance, which is very close to the bounds.         

\section{Conclusions}

%By imposing proper priors, UAMP-MF can be used to solve various problems, such as RPCA, DL, CSMU, NMF, and sparse MF. Extensive numerical results demonstrate the superiority of
%UAMP-MF in computational complexity, recovery accuracy, and robustness compared to state-of-the-art algorithms. \rew{Future work includes the use of powerful VI techniques \cite{VIUID}, \cite{liustein} to deal with more complex priors and models involved in MF.}

In this paper, we tackled the blind joint near-field localization and signal detection problem in an ISAC system, which is formulated as a MF problem with proper structures imposed on the factor matrices. By incorporating UAMP into variational inference through a whitening process, we designed the message passing algorithm UAMP-MF to solve a generic MF problem. Then, we apply the UAMP-MF algorithm to solve the joint localization and signal detection problem, where the factor matrix structures are fully exploited. Extensive simulation results demonstrate that the proposed algorithm significantly outperforms existing algorithms.   

\begin{appendices}

\section{Proof of Proposition 1}

%\begin{proof}
	According to VI, the message from $f_Y$ to $\bX$ can be expressed as 	
	\begin{eqnarray}
		&&\!\!\!\!\!\!\!\!\!\!m_{f_Y\to X}(\bX) \nonumber\\
		&&\!\!\!\!\!\!\!\!\!\!\propto \exp\left(\int_{\bA, \lambda} q(\bA)q(\lambda) \log f_Y   \right) \nonumber\\
		&&\!\!\!\!\!\!\!\!\!\!\propto \exp\Big(-\hat\lambda\int_{\bA} \Tr\Big( (\bY-\bA\bX)\herm (\bY-\bA\bX)\Big) q(\bA) \Big)\nonumber\\
%\end{eqnarray}
%\begin{eqnarray}
		&&\!\!\!\!\!\!\!\!\!\!\propto \exp\left(-\hat\lambda \int_{\bA}(\by-\tVec(\bA\bX))\herm (\by-\tVec(\bA\bX)) q(\bA) \right)\nonumber\\
		&&\!\!\!\!\!\!\!\!\!\!= \exp\left(-\hat\lambda \int_{\ba}(\by-\tilde\bX\ba))\herm (\by-\tilde\bX\ba)) q(\ba) \right)\label{eq:msg_fyX_proof1}\\
		&&\!\!\!\!\!\!\!\!\!\!=\exp\Big(-\hat\lambda \int_{\ba}\big(\by\herm \by+\underbrace{\ba\herm\tilde\bX\herm\tilde\bX\ba}_{(i)}   -\underbrace{\ba\herm\tilde\bX\herm\by}_{(ii)} -\underbrace{\by\herm\tilde\bX\ba}_{(iii)}\big) q(\ba)\Big), \nonumber \\ \label{eq:msg_fyX_proof}
	\end{eqnarray}
where
	%\begin{equation}
	%	\hat\lambda = \int_{\lambda} \lambda q(\lambda)/\int_{\lambda} q(\lambda)\nonumber
	%\end{equation}
	%with its computation shown in \eqref{eq:lambda},
$\ba=\text{Vec}(\bA)$, $\by=\text{Vec}(\bY)$ and $\tilde\bX=\bX\herm\otimes \bI_M$. In the derivation of \eqref{eq:msg_fyX_proof1}, we use the matrix identity $\text{Vec}(\bA\bX) =\text{Vec}(\bI_M\bA\bX)= \left(\bX\herm\otimes \bI_M\right) \ba$ \cite{matrixcook}. Next, we work out the integration of the three terms $(i)$, $(ii)$ and $(iii)$ in \eqref{eq:msg_fyX_proof}. For term $(i)$, we have %where the integration in the exponent part can be divided into the following there parts:
	\begin{eqnarray}
		&&\!\!\!\!\!\!\!\!\!\int_{\ba}\ba\herm\tilde\bX\herm\tilde\bX\ba q(\ba)\nonumber\\
		&&\!\!\!\!\!\!\!\!\!=\int_{\ba}\Tr\left(\tilde\bX\herm\tilde\bX\ba\ba\herm \right) q(\ba)\nonumber\\
		&&\!\!\!\!\!\!\!\!\!=\Tr\left(\tilde\bX\herm\tilde\bX(\hat\ba\hat\ba\herm+\bV_A\otimes\bU_A)\right)\nonumber\\
		&&\!\!\!\!\!\!\!\!\!=\Tr(\hat\ba\herm\tilde\bX\herm\tilde\bX\hat\ba)
		+\Tr(\tilde\bX(\bV_A\otimes\bU_A)\tilde\bX\herm)\nonumber \\
		&&\!\!\!\!\!\!\!\!\!=\Tr(\tVec(\hat\bA)\herm(\bX\herm\otimes\bI_M)\herm(\bX\herm\otimes\bI_M)\tVec(\hat\bA))\nonumber \\
		&&\ \ \ \ \ \ \ +\Tr((\bX\herm\otimes \bI_M)(\bV_A\otimes\bU_A)(\bX\herm\otimes \bI_M)) \\
		&&\!\!\!\!\!\!\!\!\!=\Tr\big(\tVec(\hat\bA\bX)\herm \tVec(\hat\bA\bX)\big) \\  %\nonumber
%\end{eqnarray}
%\begin{eqnarray}
		&&\ \ \ \ \ \ \ \ +\Tr\big((\bX\otimes \bI_M)(\bX\herm\otimes \bI_M)(\bV_A\otimes\bU_A)\big)\label{eq:msg_fyX_parti_1}\\
		&&\!\!\!\!\!\!\!\!\!=\Tr(\bX\herm\hat\bA\herm\hat\bA\bX)\nonumber\\
		&&\ \ \ \ \ \ \ \ +\Tr(((\bX\bX\herm)\otimes\bI_M)(\bV_A\otimes \bU_A))\label{eq:msg_fyX_parti_2}\\
		&&\!\!\!\!\!\!\!\!\!=\Tr(\bX\herm\hat\bA\herm\hat\bA\bX)+\Tr(\bU_A)\Tr(\bX\bX\herm\bV_A)
		\label{eq:msg_fyX_parti_3}\\
		&&\!\!\!\!\!\!\!\!\!=\Tr(\bX\herm(\hat\bA\herm\hat\bA+\Tr(\bU_A)\bV_A)\bX),  \label{eq:msg_fyX_parti}
	\end{eqnarray}
	where we use the matrix identity $\Tr\left(\bA\bB\bC\right)=\Tr\left(\bC\bA\bB\right)$ in deriving \eqref{eq:msg_fyX_parti_1},  $\left(\bA\otimes\bB\right)\left(\bC\otimes\bD\right)=(\bA\bB)\otimes(\bC\bD)$ and $\left(\bA\otimes \bB\right)\herm=\bA\herm\otimes\bB\herm$ in deriving \eqref{eq:msg_fyX_parti_2}, and $\Tr(\bA\otimes\bB)=\Tr(\bA)\Tr(\bB)$ in deriving \eqref{eq:msg_fyX_parti_3}.
	%the computation of $\hat\bA$ is shown in \eqref{eq:pseudoH3}.

	Regarding term $(ii)$, we have
	\begin{eqnarray}
		&&\int_{\ba}\ba\herm\tilde\bX\herm\by q(\ba)\nonumber\\
		&&=\int_{\bA}\text{Vec}(\bA)\herm(\bX\herm\otimes \bI_M)\herm\text{Vec}(\bY)q(\bA)\nonumber\\
		&&=\int_{\bA}\Tr(\bX\herm\bA\herm\bY)q(\bA) \nonumber\\
		&&=\Tr\big(\bX\herm\hat\bA\herm\bY\big). \label{eq:msg_fyX_partii}
	\end{eqnarray}
	Similarly,  term $(iii)$ can be expressed as
	\begin{eqnarray}
		\int_{\ba}\by\herm\tilde\bX\ba q(\ba)=\Tr(\bY\herm\hat\bA\bX). \label{eq:msg_fyX_partiii}
	\end{eqnarray}
Based on the above results, the message
	\begin{eqnarray}
		&&\!\!\!\!\!\!\!\!\!\!\!\!\!\!\!m_{f_Y\to X}(\bX) \propto \exp\Big(-\hat\lambda \Tr\big(\bX\herm(\hat\bA\herm\hat\bA+\Tr(\bU_A)\bV_A)\bX\nonumber\\
		&&\ \ \ \ \ \ \ \ \ \ \ \ \ \ \ \ \ \ -\bX\herm\hat\bA\herm\bY-\bY\herm\hat\bA\bX+\bY\herm\bY\big) \Big). % \label{eq:msg_fyX1}
		%	&&=\exp\Big(-\hat\lambda \Tr\big(\bX\herm\vec\bU_X^{-1}\bX
		%	-\bX\herm\hat\bA\herm\bY\nonumber\\
		%	&&\ \ \ \ \ \ \ \ \ \ \ \ \ \ \ \ \ \ -\bY\herm\hat\bA\bX+\bY\herm\bY\big) \Big),
	\end{eqnarray}
	%\begin{eqnarray}
	%	&&\hat\lambda\int_{\ba}\big(\by\herm \by+\ba\herm\tilde\bX\herm\tilde\bX\ba-
	%	\ba\herm\tilde\bX\herm\by -\by\herm\tilde\bX\ba\big) b(\ba)\nonumber\\
	%	&&=\Tr\big(\hat\lambda\bY\herm\bY+\bX\herm\vec\bU_X^{-1}\bX-\hat\lambda\bY\herm\hat\bA\bX-
	%	\hat\lambda\hat\bX\herm\hat\bA\herm\bY\big)\nonumber\\
	%	\label{eq:appendix1}
	%\end{eqnarray}
	%where
	%\begin{equation}
	%	\vec\bU_X^{-1}= \hat\lambda\big(\hat\bA\herm\hat\bA+\Tr(\bU_A)\bV_A\big).
	%\end{equation}
	%where $\by=\tVec(\bY)$, $\ba=\tVec(\bA)$, $\tilde\bX=\bX\herm\otimes \bI_L$ and $\vec\bU_X^{-1} = \hat\lambda\big(\hat\bA\herm\hat\bA+\Tr(\bU_A)\bV_A\big)$. More details about the derivation of \eqref{eq:msg_fyX1} is shown in Appendix A.
	Comparing the result against the matrix \rew{Gaussian} distribution, we have the result shown in \eqref{eq:msg_fyX_result}. %-\eqref{eq:msg_fyX_U}.
	%\begin{eqnarray}
	%m_{f_Y\to X}(\bX)&\propto&\MN(\bX;\vec \bX ,\vec\bU_X,\bI_L)
	%&\triangleq& \MN(\bX;\vec\bX,\vec\bU_X,\bI_L)
	%\label{eq:msg_fyX2}
	%\end{eqnarray}
	%where $\vec\bX = \hat\lambda\vec\bU_X\hat\bA\herm\bY$.
%\end{proof}

\section{Proof of Proposition 2}

%\begin{proof}
	%Similar to the computation of the message $m_{f_Y \to X}(\bX)$,
	According to VI, the message $m_{f_Y \to H}(\bA)$ is computed as
	\begin{eqnarray}
		&&m_{f_Y\to H}(\bA) \nonumber \\
		&&\propto \exp\left(\int_{\bX, \lambda} q(\bX) q(\lambda) \log f_Y \right)\nonumber\\
		&&\propto\exp\left(-\hat\lambda\int_{\bX} \Tr( (\bY-\bA\bX)\herm (\bY-\bA\bX)) b(\bX) \right)\nonumber\\
		&&=\exp\left(-\hat\lambda \int_{\bx} (\by-\tilde\bA\bx)\herm (\by-\tilde\bA\bx)b(\bx)\right)\nonumber\\
		&&=\exp\Big(-\hat\lambda \int_{\bx} \big(\by\herm \by+ \bx\herm\tilde\bA\herm\tilde\bA\bx\nonumber\\
		&&\ \ \ \ \ \ \ \ \ \ \ \ \ \ \ \ \ \ \ \ \ \ \  \ \ \ \  -\bx\herm\tilde\bA\herm\by  -\by\herm\tilde\bA\bx \big) b(\bx) \Big), \label{eq:msg_fyH_proof}
	\end{eqnarray}
	where {$\tilde\bA\triangleq \bI_L\otimes \bA$}. Similar to the derivation of \eqref{eq:msg_fyX_parti}, the integration of the terms in \eqref{eq:msg_fyH_proof} can be expressed as
	\begin{eqnarray}
		\int_{\bx}\bx\herm\tilde\bA\herm\by b(\bx)= \Tr\big(\hat\bX\herm\bA\herm\bY\big)
	\end{eqnarray}
	and
	\begin{eqnarray}
		&&\int_{\bx}\bx\herm\tilde\bA\herm\tilde\bA\bx b(\bx) \nonumber \\
		&&= \int_{\bx} \Tr\left(\tilde\bA\herm\tilde\bA\bx\bx\herm\right)b(\bx)\nonumber
	\end{eqnarray}
    \begin{eqnarray}
		&& =\Tr\big(\tilde\bA\herm\tilde\bA(\hat\bx\hat\bx\herm+\bV_X\otimes \bU_X)\big)\nonumber\\
		&&=\Tr\big(\hat\bX\herm\bA\herm\bA\hat\bX\big)+\Tr\big(\bV_X\otimes(\bA\herm\bA\bU_X)\big)\nonumber\\
		&&=\Tr\big(\hat\bX\herm\bA\herm\bA\hat\bX\big)+\Tr(\bV_X)\Tr\big(\bA\bU_X\bA\herm\big)\nonumber\\
		&&=\Tr\big(\bA(\hat\bX\hat\bX\herm+\Tr(\bV_X)\bU_X)\bA\herm\big).
	\end{eqnarray}
The message from $f_Y$ to $\bA$ can be represented as
	\begin{eqnarray}
		&&m_{f_Y\to H}(\bA)\nonumber\\
		&&\propto\exp\Big(-\hat\lambda \Tr(\bY\bY\herm+\bA(\hat\bX\hat\bX\herm+\Tr(\bV_X)\bU_X)\bA\herm\nonumber\\
		&&\ \ \ \ \ \ \ \ -\bA\hat\bX\bY\herm-\bY\hat\bX\herm\bA\herm) \Big).
	\end{eqnarray}
Comparing the above against the matrix \rew{Gaussian} distribution, we obtain the result shown by \eqref{eq:msg_fyH_result} - \eqref{eq:msg_fyH_V}.
	%$m_{f_Y\to H}(\bA)\propto \MN\left(\bA; \cev\bA, \bI_M, \cev\bV_A\right)$ with $\cev\bV_A = \hat\lambda^{-1}\left(\hat\bX\hat\bX\herm+\Tr(\bV_X)\bU_X\right)^{-1}$ and
	%$\cev\bA= \hat\lambda\bY\hat\bX\herm\cev\bV_A\label{eq:msg_fyH}$.
%\end{proof}

\section{Proof of Proposition 3}
%\begin{proof}
	According to VI, the message
	\begin{eqnarray}
		&&m_{f_Y\to \lambda}(\lambda) \nonumber \\
		&& =\text{det}(\lambda^{-1}\bI_M\otimes\bI_L)\exp\Big(-\lambda \big(\text{Vec}
		(\bY)-\tVec(\bA\bX)\big)\herm \nonumber\\
		&&\ \ \ \ \ \ \ \ \ \ \ \ \ \ \ \ \ \ \ \ \ \ \ \ \ \ \ \ \ \ \ \ \ \ \ \ \   \big(\tVec(\bY)-\tVec(\bA\bX)\big)\Big)\nonumber\\
		&&=\lambda^{ML}\exp\Big(-\lambda\int_{\bA,\bX}\Tr((\bY-\bA\bX)\herm\nonumber\\
		&&\ \ \ \ \ \ \ \ \ \ \ \ \ \ \ \ \ \ \ \ \ \ \ \ \ \ \  \ \ \ \ \ \ \  (\bY-\bA\bX))b(\bA)b(\bX)\Big)\nonumber\\
		&&=\lambda^{ML}\exp\Big(-\lambda C \Big),
	\end{eqnarray}
	where
	\begin{eqnarray}
		&&\!\!\!\!\!\!\!\!\! C=\int_{\bA,\bX}
		\Tr((\bY-\bA\bX)\herm(\bY-\bA\bX))q(\bA)q(\bX)\nonumber\\
		&&\!\!\!\!\!\!\!\!\! \!=\!\int_{\bx,\bA} (\by\herm\by+\bx\herm\tilde\bA\herm\tilde\bA\bx-\bx\herm\tilde\bA\herm\by-\by\herm\tilde\bA\bx)
		q(\bx)q(\bA)\nonumber\\
		&&\!\!\!\!\!\!\!\!\! \!=\!\int_{\bA} \Tr(\bY\bY\herm+\bA(\hat\bX\hat\bX\herm+\Tr(\bV_X)\bU_X)\bA\herm\nonumber\\
		&&\ \ \ \ \ \ \ \ \ \ \ \ \ \ \ \ \ \ \ \ \ \ \ \ \ \ \ \  -\bA\hat\bX\bY\herm-\bY\hat\bX\herm\bA\herm)q(\bA)\nonumber\\
		&&\!\!\!\!\!\!\!\!\!\! =\Tr(\bY\bY\herm -\hat\bA\hat\bX\bY\herm-\bY\hat\bX\herm\hat\bA\herm)\nonumber\\
		&& +\Tr((\hat\bX\hat\bX\herm+\Tr(\bV_X)\bU_X)(\hat\bA\herm\hat\bA+\Tr(\bU_A)\bV_A)\herm)\nonumber\\
		&&\!\!\!\!\!\!\!\!\!\! =\Tr\Big(\big(\bY-\hat\bA\hat\bX\big)\herm\big(\bY-\hat\bA\hat\bX\big)\Big)
		+\Tr\Big(\hat\bX\hat\bX\herm\Tr(\bU_A)\bV_A\nonumber\\
		&&+\Tr(\bV_X)\bU_X\hat\bA\herm\hat\bA+\Tr(\bV_X)\bU_X\Tr(\bU_A)\bV_A\Big).
	\end{eqnarray}
	By simplifying the above result, we obtain \eqref{eq:ComputeC}.
	%Derivation of the integration in the exponent part is shown in Appendix B.
%\end{proof}

\end{appendices}

%\section*{Acknowledgement}
%The authors would like to thank Dr. Kejun Huang for sharing the code for AO-ADMM.

\bibliographystyle{IEEEtran}
\bibliography{bibliography}

\end{document}